\documentclass[aps, prx, twocolumn]{revtex4-2}

\usepackage{graphicx}
\usepackage{epstopdf}
\usepackage{dcolumn}   
\usepackage{amsmath,amssymb,dsfont}
\usepackage{wrapfig}
\usepackage{array}
\usepackage[caption=false]{subfig}
\usepackage[bookmarks=false,linkcolor=blue,urlcolor=blue,colorlinks,citecolor=blue]{hyperref}
\usepackage{exscale,relsize}
\usepackage[usenames, dvipsnames]{color}
\usepackage{bbold}
\usepackage{xcolor}

\usepackage[]{cleveref}
\Crefformat{appendix}{Appendix~#2#1#3}
\Crefname{section}{Sec.}{Secs.}
\Crefname{subsection}{Sec.}{Secs.}
\Crefname{appendix}{\IfAppendix{Sec.}{Sec.}}{\IfAppendix{Secs.}{Secs.}}
\Crefname{subappendix}{\IfAppendix{Sec.}{Sec.}}{\IfAppendix{Secs.}{Secs.}}
\Crefname{equation}{Eq.}{Eqs.}
\Crefname{figure}{Fig.}{Figs.}
\Crefname{tabular}{Tab.}{Tabs.}

\usepackage{microtype}

\usepackage{siunitx}
\DeclareSIUnit{\ueV}{\micro\electronvolt}
\DeclareSIUnit{\um}{\micro\meter}
\DeclareSIUnit{\nm}{\nano\meter}
\DeclareMathOperator{\erf}{erf}
\DeclareMathOperator{\SNR}{SNR}

\newcommand{\Q}{\mathrm{\scriptscriptstyle Q}}
\newcommand{\Cq}{C_\Q}
\newcommand{\dCq}{\Delta \Cq}

\newcommand{\X}{{\scriptscriptstyle X}}
\newcommand{\Z}{{\scriptscriptstyle Z}}
\newcommand{\tauX}{\tau_\X}
\newcommand{\tauXup}{\tau_{\X\uparrow}}
\newcommand{\tauXdown}{\tau_{\X\downarrow}}
\newcommand{\tauXud}{\tau_{\X\uparrow/\downarrow}}
\newcommand{\tauZ}{\tau_\Z}
\newcommand{\tauZup}{\tau_{\Z\uparrow}}
\newcommand{\tauZdown}{\tau_{\Z\downarrow}}

\newcommand{\err}{\mathsf{err}}
\newcommand{\erra}{\err_a}
\newcommand{\errXa}{\erra^\X}
\newcommand{\errZa}{\erra^\Z}

\newcommand{\Ed}{E_\mathrm{\scriptscriptstyle D}}

\newcommand{\M}{{\scriptscriptstyle M}}

\newcommand{\RO}{\mathrm{\scriptscriptstyle RO}}
\newcommand{\RTS}{\mathrm{\scriptscriptstyle RTS}}

\newcommand{\Bpara}{B_\parallel}
\newcommand{\Bperp}{B_\perp}

\newcommand{\VQDL}{V_\mathrm{QDL}}
\newcommand{\VQD}{V_\mathrm{QD1}}

\newcommand{\bra}[1]{\langle #1|}
\newcommand{\ket}[1]{| #1\rangle}

\newcommand{\QD}[1]{\text{QD}_{#1}}

\newcommand{\tm}[1]{t_{\text{m}#1}}

\newcommand{\kB}{k_\mathrm{\scriptscriptstyle B}}

\begin{document}

\title{Distinct Lifetimes for $X$ and $Z$ Loop Measurements in a Majorana Tetron Device}

\author{Microsoft Quantum$^\dagger$}
\noaffiliation
\date{\today}

\begin{abstract}
We present a hardware realization and measurements of a tetron qubit device in a superconductor-semiconductor heterostructure.
The device architecture contains two parallel superconducting nanowires, which support four Majorana zero modes (MZMs) when tuned into the topological phase, and a trivial superconducting backbone.
Two distinct readout interferometers are formed by connecting the superconducting structure to a series of quantum dots.
We perform single-shot interferometric measurements of the fermion parity for the two loops, designed to implement Pauli-$X$ and $Z$ measurements of the tetron.
Performing repeated single-shot measurements yields two widely separated time scales $\tauX = \SI{14.5\pm 0.3}{\micro\second}$ and $\tauZ = \SI{12.4\pm 0.4}{\milli\second}$ for parity switches observed in the $X$ and $Z$ measurement loops, which we attribute to intra-wire parity switches and external quasiparticle poisoning, respectively.
We estimate assignment errors of $\errXa = 16\%$ and $\errZa = 0.5\%$ for $X$ and $Z$ measurement-based operations, respectively.
\end{abstract}

\maketitle

\paragraph{Introduction.} \label{sec:intro}

The use of measurement-based topological qubit arrays for quantum computation is attractive for several reasons: 
it is predicted that most physical qubit errors can be suppressed exponentially in several dimensionless quantities~\cite{Kitaev97, Freedman98, Nayak08}; the requisite topological phase can occur in superconductor-semiconductor hybrid nanowires that are amenable to scalable fabrication techniques~\cite{Kitaev01, Lutchyn10, Oreg10, Suominen17}; 
qubit control is ``digital'' in the sense that error rates do not depend sensitively on the amplitude of the baseband control pulses, simplifying calibration and reducing susceptibility to noise; 
operations can be performed rapidly, on timescales~\cite{Plugge17, Karzig17} expected to allow utility-scale applications to be executed in weeks~\cite{Beverland22}; 
and the qubit footprint is small enough to permit millions of qubits on a single chip~\cite{Aasen25}.
These benefits have spurred a wide range of activity~\cite{Fu09, Akhmerov09, Fu09a, Fu10, Hassler10, Sau11a, Fidkowski11b, Alicea12a, DasSarma15, Heck11, Heck12, Hyart13, Houzet13, Pientka13a, Cheng15, Sau15, Yavilberg15, Aasen16, Hell18, Chiu18, Drukier18, Vijay15, Vijay16a, Vijay16b, Knapp18a, Knapp18b, Liu19, Munk20, Steiner20, Khindanov21a}, including a recent roadmap to a fault-tolerant quantum computer based on topological qubits~\cite{Aasen25}. 

\begin{figure}
\includegraphics[width=\columnwidth]{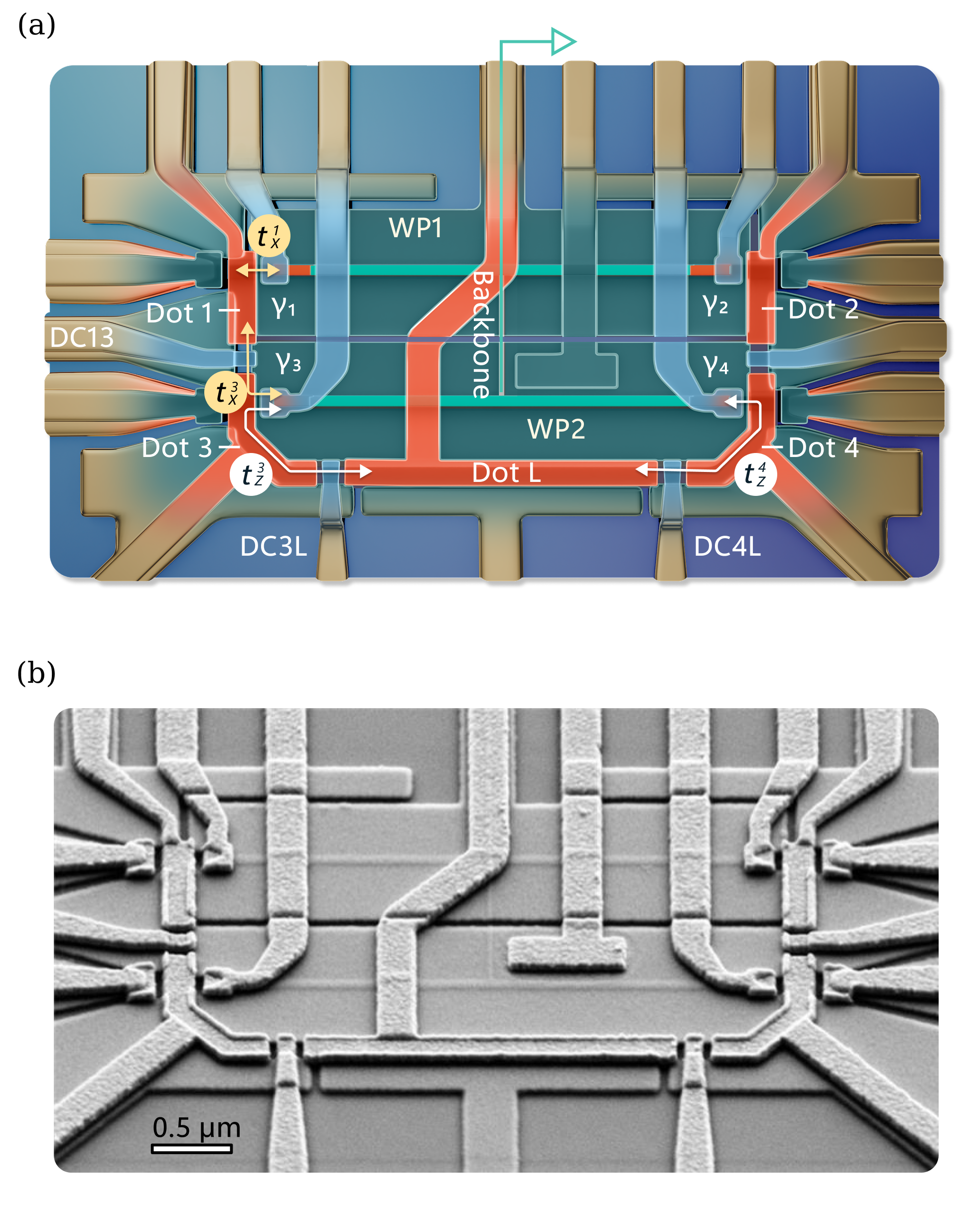}
\caption{
(a)~Schematic of a tetron, highlighting the hybrid nanowires, backbone, quantum dots, and junctions. 
The H-shaped aluminum pattern (teal) defines the two topological nanowires (horizontal) and the trivial superconducting backbone (vertical) which has an extension that is connected to ground. 
Wire plunger gates WP1 and WP2 are used to tune the horizontal wire segments into the lowest electric subband while simultaneously fully depleting the narrower backbone.
In the topological phase, there are MZMs at the four ends of the H, highlighted in brown and labeled $\gamma_i$. 
The high-SNR readout dots 1 and L are coupled to the wires via the indicated
tunnel couplings $t_\X^{1,3}$, $t_\Z^{3,4}$. 
These tunnel couplings are controlled by cutter gates DC13, DC3L, DC4L (light blue) and the detuning of the quantum dots which is set by the dot plunger gates (orange). 
Gold-colored gates on the periphery are used to establish confining potentials for the quantum dots and transport paths.
(b)~An SEM image of a tetron.
 }
\label{fig:tetron}
\end{figure}

The first step in that roadmap is the construction of a so-called tetron device comprised of two topological nanowires~\cite{Kitaev01, Lutchyn10, Oreg10} connected by a trivial superconducting wire~\cite{Karzig17, Plugge17} (see \Cref{fig:tetron}(a)). 
The tetron is designed to support four Majorana zero modes (MZMs), which encode a single qubit when the total parity is fixed; the parity eigenstates of either topological wire are a convenient set of basis states. 

One of the basic operations in a tetron is a measurement of the fermion parity of a topological nanowire~\cite{Aghaee25}. 
This is accomplished by forming an interference loop~\cite{Plugge16, Plugge17, Karzig17} comprised of one of the topological nanowires and a quantum dot (referred to hereafter as a $Z$ loop).
(Alternatively, it can
be accomplished by increasing MZM hybridization
through the wire \cite{vanLoo25}.)
The quantum capacitance $\Cq$ of the dot is a function of the total parity of the dot and the two MZMs in the loop. 
Probing it via dispersive gate-sensing~\cite{Colless13} constitutes a Pauli-$Z$ measurement of a tetron.
To perform a Pauli-$X$ measurement in this encoding, one similarly measures the fermion parity along an interference loop that bridges the two nanowires and thus involves Majorana zero modes (MZMs) located on distinct nanowires (referred to hereafter as an $X$ loop).

Here, we report on measurements of a device realizing the tetron concept. 
We present evidence that the two distinct types of measurements that we perform probe the fermion parity on $Z$ and $X$ loops of the tetron. 
The minimum observed single-shot measurement-errors are $0.5\%$ for the $Z$ loop and $16\%$ for the $X$ loop. 
Continuous monitoring of the two measurements reveals distinct characteristic timescales of $\tauZ = \SI{12.4 \pm 0.4}{\milli\second}$ and $\tauX = \SI{14.5 \pm 0.3}{\micro\second}$. 
We interpret these as fermion parity switches and show that they are consistent with a theoretical model in which $\tauZ$ is set by quasiparticle poisoning and $\tauX$ is set by thermal excitations and quantum jumps due to residual splitting between low energy modes localized at the ends of the nanowires.

We argue that these results are the first demonstration of two distinct projective measurements of fermion parity in a tetron device. 
When performed in a topological qubit, these measurements correspond to Pauli measurements in orthogonal bases~\cite{Bonderson08a, Bonderson08b}. Confirming this orthogonality using a rapid succession of $X$ and $Z$ loop measurements is the next step of our roadmap~\cite{Aasen25}.
These Pauli measurements are the only single-qubit operations required to perform quantum error correction and can be augmented with non-Clifford operations such as $T$ gates to achieve universality.

\paragraph{Device overview.}

Our device has been fabricated using a high-quality
\footnote{We observe mobilities typically in excess of \SI{60000}{\centi\meter^2/\volt\second} and, in some growths, exceeding \SI{100000}{\centi\meter^2/\volt\second}.}
InAs two-dimensional electron gas (2DEG) grown by molecular beam epitaxy and a two-metal-layer fabrication process, similar to the method described in Ref.~\onlinecite{Aghaee25}. 
(See \Cref{app:design} for more information and for a complete device schematic and gate naming convention; throughout the manuscript, $V_i$ refers to the dc voltage applied to gate $i$.) 
Two $\SI{3.5}{\micro\meter}$-long Al strips are connected by a narrower $\SI{1}{\micro\meter}$-long superconducting backbone, forming a sideways ``H,'' shown in teal in \Cref{fig:tetron}(a).
Each Al strip is covered by a plunger gate, denoted as WP1 (WP2) for the top (bottom) wire, highlighted in green in \Cref{fig:tetron}(a).
The plunger gates deplete the 2DEG, thereby forming the semiconductor nanowires, and enable tuning the proximitized semiconducting nanowires into the single-subband regime. 
Because the backbone is narrower, these plunger gates simultaneously fully deplete the 2DEG underneath it. 
To facilitate device tuning through non-local conductance measurements, as used in the topological gap protocol (TGP)~\cite{Pikulin21, Aghaee23}, the superconductor is connected to ground through an extension of the backbone which runs off the top of the image in \Cref{fig:tetron}(a). 
The ends of the tetron device are connected to the dots via tunnel junctions to form interferometer loops. 
A scanning electron micrograph of a nominally identical device is shown in \Cref{fig:tetron}(b).
This design facilitates more efficient multi-qubit layouts~\cite{Karzig17}
than the linear design of Ref.~\onlinecite{Aghaee25}.

\begin{figure*}
\centering
\includegraphics[width=0.85\paperwidth]{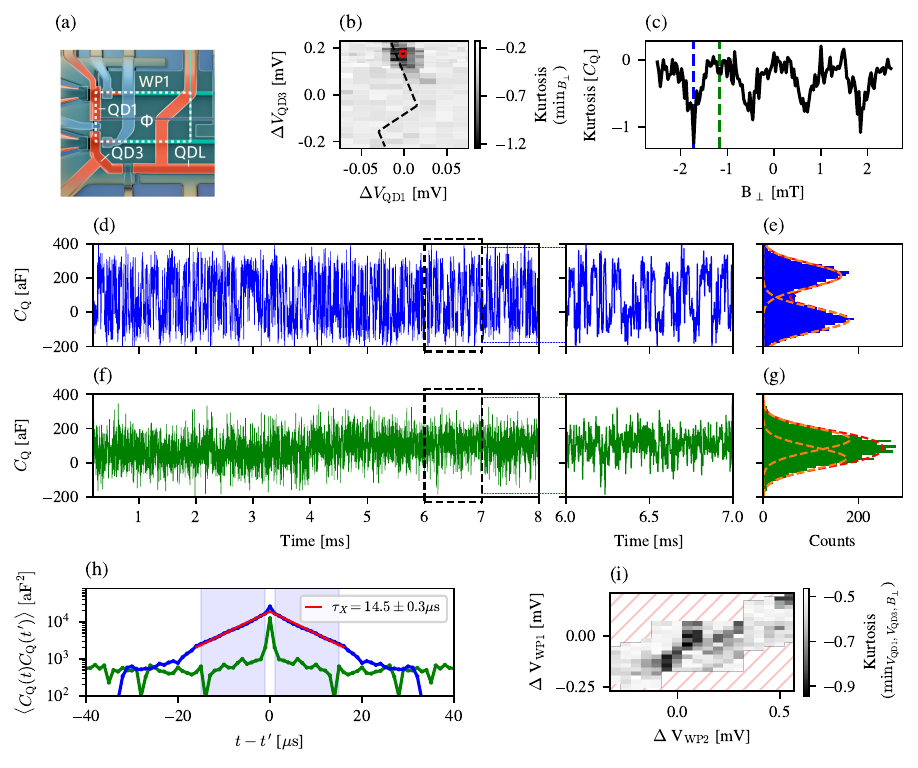}
\caption{
(a)~Schematic of the interference loop involved in the $X$ measurement. Gate coloring indicates the associated tuning as described in \Cref{fig:tetron_gates}; $\Phi$ denotes the magnetic flux threading the loop, controlled by $\Bperp$. 
(b)~Excess kurtosis of measured $\Cq$ time record as a function of varations of $V_\mathrm{QD1}$ and $V_\mathrm{QD3}$ near charge degeneracy.
The red circle denotes the gate voltages used in panels (c)-(h). The dashed black line highlights the position of interdot and dot-wire charge transitions, see \Cref{fig:average_qd_qd} for details. 
(c)~Excess kurtosis of measured $\Cq$ time record as a function of $\Bperp$ at fixed gate voltages.
(d)~Measured $\Cq$ time record at the flux value indicated by a dashed blue line in panel~(c) using an effective integration time of $\SI{2}{\micro\second}$ per data point. The inset to the right shows a zoom-in of the data marked by the black dashed box to exemplify the observed RTS.
(e)~A histogram of measured $\Cq$ values over the full time record
    in panel (d).
(f,g) Same quantities as in (d,e) at the flux value indicated by a dashed green line in panel~(c).
(h)~Autocorrelation functions computed from the time records in panels~(d) and (f). The blue data points associated with the former show a clear exponential decay over time differences up to $\SI{30}{\micro\second}$, indicative of a RTS. 
The red line is a fit to the data in the shaded region, which results in $\tauX = \SI{14.5 \pm 0.3}{\micro\second}$.
(i)~The excess kurtosis as a function of the difference in $V_{\mathrm{WP1}}$ and $V_{\mathrm{WP2}}$ relative to the value used in the initial measurement shown in (c-h). 
The dark region near the middle highlights the optimal tuning point which shifted slightly in the intervening days. 
Red cross-hatching indicates regions in voltage space where no data were collected.
} 
\label{fig:xmpr}
\end{figure*}

\begin{figure}
\centering
\includegraphics[width=\columnwidth]{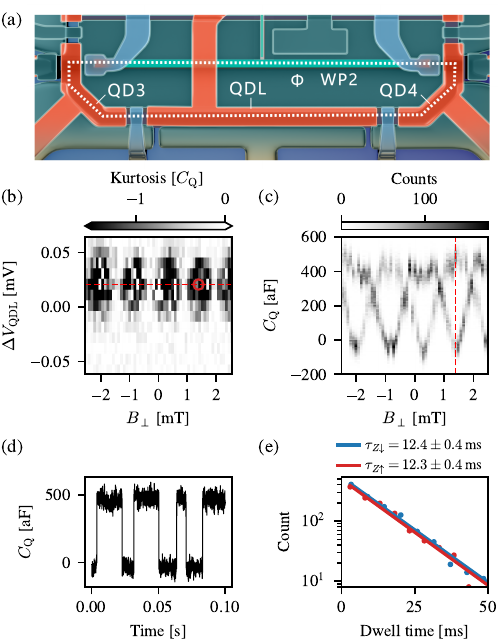}
\caption{
(a)~The interference loop involved in the $Z$ measurement, enclosing flux $\Phi$. 
Gate coloring indicates the associated tuning as described in \Cref{fig:tetron_gates}.
(b)~Excess kurtosis of the quantum capacitance measured on QDL as a function of out-of-plane field and $\VQDL$ when the system is tuned into the loop configuration appropriate for the $Z$ measurement.
The voltage $V_\mathrm{QD3}$ ($V_\mathrm{QD4}$) is kept fixed at a value corresponding to a detuning of $\sim 0.1$ $(0.1)$ electrons from charge resonance.
We observe dips in the excess kurtosis, indicating the presence of a RTS, with a periodicity of $\SI{1.1}{\milli\tesla}$ near charge degeneracy of the quantum dot.
(c)~Histogram of the $\Cq$ record as a function of $\Bperp$ for the $V_\mathrm{QDL}$ detuning value indicated by the red dashed line in (b).
(d)~Example time record corresponding to the red marker in (b) and vertical dashed line in (c).
(e)~Dwell time distribution of the two states of the RTS extracted from all time records in the data (sweeping $\VQDL$, $V_\mathrm{QD4}$ and $B_{\perp}$) for which the excess kurtosis $<-1$. Solid lines correspond to an exponential fit with an average dwell time of $\tau_\Z = \SI{12.4 \pm 0.4}{\milli\second}$~\cite{Ztimescale_footnote}. 
}
\label{fig:zmpr}
\end{figure}

The readout interferometers (two of which are used in this paper) consist of five semiconductor quantum dots: dots 1-4  are adjacent to the ends of the nanowires and dot L is parallel to the bottom nanowire, as shown in \Cref{fig:tetron}(a). 
Two loops are discussed in this paper.
The $X$ loop is formed using dot 1, dot 3, the left halves of the
two nanowires, and the backbone, as shown in \Cref{fig:xmpr}(a). 
The $Z$ loop is formed using dots 3, L, and 4, and the
bottom nanowire, as shown in \Cref{fig:zmpr}(a).
Measurements of the quantum capacitance of dots 1 and L will be referred to as ``$X$ loop measurements'' and ``$Z$ loop measurements,'' respectively, when their corresponding loops are connected.

The magnetic flux through these loops is controlled with a perpendicular magnetic field $\Bperp$. 
When the wires are in the topological phase, the dots are coupled to Majorana zero modes ($\gamma_i$ in \Cref{fig:tetron}(a)), and their quantum capacitances $\Cq$ have flux-dependent sensitivity to the joint parity of the distinct pairs of MZMs coupled by the corresponding loops~\cite{Aghaee25, Karzig17, Boutin25}.
Under these conditions, the capacitance difference $\dCq$ between the two parities is $h/2e$-periodic, as discussed in Ref.~\onlinecite{Aghaee25} for the $Z$-loop and in \Cref{app:model} for the $X$ loop.

To maximize the visibility of $\dCq$, the interferometer
loops need to be balanced.
In the topological phase, the $X$ loop involves effective couplings $t_\X^{1,3}$ as indicated in \Cref{fig:tetron}(a), where $t_\X^{(1)}$ is a direct coupling of dot 1 to $\gamma_1$, while $t_\X^{(3)}$ is an indirect coupling to $\gamma_3$ controlled by DC13, dot 3, and its coupling to the bottom nanowire.
The analogous $Z$ loop effective couplings from dot L to $\gamma_{1,3}$, are $t_\Z^{3,4}$, which are controlled by DC3L, DC4L, dot 3, dot 4, and their respective couplings to the nanowire~%
\footnote{In principle, one can also define loops corresponding to a $Y$ measurement, e.g. using dots 1, 3, L and 4. However, this would require suppressing the coupling between dot 3 and its adjacent zero mode. We will not use such loops in this paper.}.
\Cref{fig:tetron_gates} contains a device layout with a complete labeling of the gates and their operational voltage regimes.

To measure $\Cq$, we employ a highly sensitive, high-bandwidth radio-frequency (rf) measurement using readout resonators in a reflectometry configuration. 
While following a similar general approach as Ref.~\onlinecite{Aghaee25}, we have improved several system components to achieve a four-fold improvement of sensitivity, as detailed in \Cref{app:readout_system}.

\paragraph{Experimental results.} \label{sec:measurements}

We now briefly describe the tune-up of the tetron device which follows a similar procedure to the one described in Ref.~\onlinecite{Aghaee25} and which is described in more detail in \Cref{app:device_tuning}.
The underlying rationale is to 
(1)~tune the wires into the topological phase and the backbone into the trivial superconducting phase in which the 2DEG is fully depleted; 
(2)~balance the $X$ or $Z$ loop interferometer approximately; 
(3)~measure a time record of $\Cq$ for dot 1 or L; 
(4)~vary dot detunings for maximum visibility of the interference signal; 
(5)~vary the flux through the interference loop and look for an $h/2e$-periodic bimodal signal;  
(6)~vary wire plungers to explore a region within the topological phase.

With the nanowires defined by the appropriate wire plunger (WP) gate voltages and transport paths set by the depletion gates, the TGP is performed simultaneously on both horizontal nanowire segments.  After fixing the in-plane magnetic field to $\SI{2.3}{\tesla}$, a value at which both segments pass the TGP (see \Cref{fig:tgp_phase_diagram}), we isolate the device from the leads by depleting the four source cutter gates SC1-4 and electrostatically define dots 1 and 3 to complete the $X$ loop. DC24 is set to depletion to eliminate any coupling through the right loop, and DC3L and DC4L are set to depletion to turn off coupling through the $Z$-loop [see \Cref{fig:xmpr}(a) for an overview of gate tunings as indicated by gate coloring].
To account for minor couplings between the gates used to tune the quantum dots and the nanowire chemical potential,
we verify the locations of the zero-bias peaks (ZBPs) in junctions 2 and 4 as a function of $V_\mathrm{WP1}$ and $V_\mathrm{WP2}$ using a local differential conductance measurement (see \Cref{fig:transport_x_loop}).
$V_\mathrm{WP1}$ and $V_\mathrm{WP2}$ are then set near the midpoint of the respective ZBP extents. Interdot and dot-wire couplings are extracted as described in \Cref{app:device_tuning}, where we explain how the couplings are tuned to balance the interferometer so that $t_\X^{(1)} \approx t_\X^{(3)} \approx \SI{1.5}{\micro\electronvolt}$.

With the interferometer roughly balanced, we measure time records of $\Cq$~%
\footnote{The data acquisition of the raw data corresponds to a close-to-boxcar integration of $\SI{1}{\micro\second}$ per data point with some additional filtering as discussed in \Cref{app:readout_system}.c.
The data shown in \Cref{fig:xmpr} is coarsened (except for the autocorrelation analysis) to an effective integration time of $\SI{2}{\micro\second}$.} 
while varying dot 1 and dot 3 detunings, to optimize interferometer balancing, and while varying the out-of-plane field $\Bperp$~%
\footnote{As was done in Ref.~\onlinecite{Aghaee25}, we sweep the $x$-axis of our vector magnet.
Due to mechanical offsets, this results in a negligible misalignment ($< \SI{0.25}{\deg}$) between $B_x$ and $\Bperp$, estimated from the angular dependence of local and non-local differential conductance.  
Throughout the manuscript, we use $\Bperp$ for clarity.  
We also note that $B_z$ is similarly misaligned from the nanowire axis.  In all transport measurements, such as TGP, we sweep a corrected field vector $\Bpara$ which results in a finite $B_x$ component.  
In all figures with $\Bperp$ as an axis, the $B_x$ value used in TGP for $\Bpara = \SI{2.3}{\tesla}$ is subtracted to emphasize that $\Bperp$ is swept symmetrically around $0$.}  
to identify flux dependence. 
Here, we start by examining the results at a set of $V_\mathrm{WP}$ values that best exemplifies the expected behavior of a tetron and will comment on the robustness over a range of wire plunger values below. \Cref{fig:xmpr}(b) shows a map of the excess kurtosis minimized over $\Bperp$ as a function of the two dot detunings.
The excess kurtosis uses the fourth moment to detect deviations from a Gaussian distribution (for which it vanishes); bimodal distributions, such as those arising from a random telegraph signal (RTS), exhibit negative excess kurtosis values~\cite{Aghaee25}.
We observe a minimum when dot 1 is close to degeneracy and dot 3 is detuned, as expected since the signal relies on fluctuations of the charge on dot 1 while dot 3 controls the effective coupling between dot 1 and the bottom wire.
The excess kurtosis minimum indicates that the $\Cq$ time record exhibits a non-Gaussian distribution for certain $\Bperp$. We examine this further in \Cref{fig:xmpr}(c), where we plot the excess kurtosis as a function of $\Bperp$ for a fixed value of dot 1 and dot 3 detunings, indicated by the red circle in \Cref{fig:xmpr}(b). We observe periodically spaced dips in the excess kurtosis with a field periodicity of $\SI{1.2}{\milli\tesla}$ extracted by fitting a sinusoid to the corresponding standard deviation of $\Cq$ with respect to $\Bperp$.  
This period is consistent with the expectation of $\SI{1.2}{\milli\tesla}$ for a flux periodicity of $h/2e$ given the lithographic area ($\approx \SI{1.75}{\micro\meter}^2$) of the $X$ loop. 

We now discuss the temporal fluctuations of $\Cq$. Overall, we observe a combination of two random telegraph processes occurring on distinct timescales — typically around $\SI{10}{\micro\second}$ and $\SI{10}{\milli\second}$, respectively, in the identified optimal regions. Focusing first on the faster dynamics, \Cref{fig:xmpr}(d) and (f) present time records of $\Cq$. A pronounced RTS is evident
in \Cref{fig:xmpr}(d), which corresponds to the dip in excess kurtosis indicated by the dashed blue line at $\Bperp \approx \SI{-1.7}{\milli\tesla}$ in \Cref{fig:xmpr}(c).
The corresponding histogram in \Cref{fig:xmpr}(e) displays clear bimodality, consistent with a RTS. In contrast, away from the optimal flux value, the time record shown in \Cref{fig:xmpr}(f) and its histogram in \Cref{fig:xmpr}(g) does not show clear bimodality. However, the distribution is broader than either of the individual Gaussians in \Cref{fig:xmpr}(e), and the excess kurtosis is nonzero though small. As discussed in \Cref{app:model}, we interpret this behavior as arising from a superposition of two overlapping Gaussian distributions. We expect this when the flux is near a multiple
of half of the flux quantum, which results from residual $\dCq$ caused by an MZM-splitting–induced avoided level crossing and slow global parity switching events, see \Cref{fig:spectrum_flux}. Indeed, the analysis of $\SI{120}{\milli\second}$-long time traces, as discussed in \Cref{app:qpp_x}, reveals the presence of an additional RTS occuring on a timescale of approximately $\SI{10}{\milli\second}$. We attribute these switching events to extrinsic quasiparticle poisoning, which alters the global fermion parity of the tetron-dot system.

To extract characteristic timescales of the fast fluctuations, we analyze the autocorrelation function of the quantum capacitance signal:
\begin{equation}
    \label{eqn:autocorr}
    \mathcal{A}[\Cq](t-t') = \bigl\langle \left( \Cq(t) - \langle \Cq \rangle \right) \left( \Cq(t') - \langle \Cq \rangle \right) \bigr\rangle,
\end{equation}
where $\langle \ldots \rangle$ denotes the average over the entire time record. 
The autocorrelation is expected to be governed by two primary contributions: 
(i)~readout noise, which is temporally uncorrelated and contributes to a delta-function term: $\mathcal{A}[\Cq]_\RO(t-t') = \sigma_{\Cq}^2 \delta(t-t')$, where $\sigma_{\Cq}$ is the standard deviation of the readout noise; and 
(ii)~RTS, which is exponentially-correlated in time and leads to a contribution $\mathcal{A}[\Cq]_\RTS(t-t') = (\dCq/2)^2 \exp(-2|t-t'|/\tauX)$, where $\dCq$ is the amplitude of the telegraph switches and $\tauX$ is the characteristic timescale.
The latter can be related to the characteristic switching times from the high-$\Cq$ ($\tauXdown$) and low-$\Cq$ ($\tauXup$) states  as $\tauX = 2 / (\tauXup^{-1} + \tauXdown^{-1})$. 
\Cref{fig:xmpr}(h) shows the autocorrelation function for the two time records shown above. For the trace in \Cref{fig:xmpr}(d), corresponding to the dip in excess kurtosis as a function of $\Bperp$, we observe a clear exponential decay extending over more than an order of magnitude, up to $|t - t'| \approx \SI{30}{\micro\second}$. 
Fitting the autocorrelation function in \Cref{fig:xmpr}(d) to the model
 $\mathcal{A}[\Cq] = \mathcal{A}[\Cq]_\RO + \mathcal{A}[\Cq]_\RTS$, where $\sigma_{\Cq}$, $\dCq$ and $\tauX$ are free parameters, yields a switching timescale of $\tauX = \SI{14.5 \pm 0.3}{\micro\second}$ with $\dCq = \SI{276 \pm 4}{\atto\farad}$ and $\sigma_{\Cq} = \SI{83 \pm 3}{\atto\farad}$.
In contrast, the autocorrelation for the second trace [\Cref{fig:xmpr}(h)] exhibits only a sharp peak at $t=t'$, indicating the absence of significant temporal correlations beyond readout noise within the experimental resolution.
This timescale in \Cref{fig:xmpr}(d,h) is roughly 3 times longer than in the device reported in Ref.~\onlinecite{Nayak25}, which may be due in part to improved device tuning.

The $X$ measurement described above is repeated for a grid of $V_\mathrm{WP1}$ and $V_\mathrm{WP2}$ values. To explore the extent of the observed flux-periodic bimodality, \Cref{fig:xmpr}(i) shows the excess kurtosis minimized over $\VQD$, $V_\mathrm{QD3}$, and $B_\perp$. As both wire plunger voltages are varied, we observe that the excess kurtosis remains below
$-0.5$ across several hundred
microvolts in both wire plunger voltages,
and falls below $-0.7$ over a range of
tens of microvolts.
This indicates that the non-Gaussian behavior is robust against small variations in the wire plunger voltages. We comment on these voltage ranges in the discussion at the end of this paper. Notably, the measurements shown in \Cref{fig:xmpr}(c-h) and \Cref{fig:xmpr}(i) were taken up to 5 days apart. Between these two
measurements, we observed a small drift of $\approx \SI{100}{\micro\volt}$ in the WP voltages corresponding to the minimal excess kurtosis.

We next interrupt the $X$ loop by depleting DC13 and opening DC3L and DC4L to form the $Z$ loop comprised of dots 3, L, and 4 [see \Cref{fig:zmpr}(a) for an overview of gate tunings]. 
Note that due to finite cross capacitance from DC3L and DC4L to the nanowire, we adjust $V_\mathrm{WP1}$ and $V_\mathrm{WP2}$ slightly relative to the values used in the $X$ loop measurement. We discuss this in \Cref{app:device_tuning}, where we show how the $Z$ interferometer loop is balanced using a similar procedure as for the $X$ loop.
With the $Z$ loop balanced, we measure time records of $\Cq$ as a function of dot detunings and $\Bperp$. The excess kurtosis of these time records, shown in \Cref{fig:zmpr}(b), displays evenly spaced dips with a field periodicity of $\SI{1.1}{\milli\tesla}$,
extracted from a sinusoidal fit of the standard deviation of $\Cq$
as a function of $B_\perp$. 
This period is consistent with the expectation of \SI{1.2}{\milli\tesla} for a flux periodicity of $h/2e$ given the lithographic area
($\approx \SI{1.7}{\micro\meter}^2$) of the $Z$ loop. At points of maximal visibility, we observe a random telegraph signal as shown in \Cref{fig:zmpr}(d) with a characteristic time of $\tauZ \equiv 2 / (\tauZup^{-1} + \tauZdown^{-1}) = \SI{12.4\pm0.4}{\milli\second}$ extracted from the dwell time distribution shown in \Cref{fig:zmpr}(c). We have confirmed that no faster timescale can be distinguished when we take time records at the same resolution of $\SI{1}{\micro\second}$ as for the $X$ loop.
The observed timescale is an order of magnitude longer than in Ref.~\onlinecite{Aghaee25}, possibly due to improvements in shielding and filtering and/or the difference in aluminum geometry between the two devices.

\paragraph{Minimal model.} \label{sec:theory}

The large difference in timescales $\tauX$ and $\tauZ$ clearly indicate a difference in physical origin, and we now introduce a minimal model that is consistent with these key experimental observations. The interference patterns shown in \Cref{fig:xmpr,fig:zmpr} are interpreted as arising from coherent single-electron tunneling around the respective loops. Motivated by the TGP results obtained for this device
and the challenges encountered
in fitting the $Z$ measurement to a quasi-MZM model \cite{Aghaee25}, we model the system using an effective Hamiltonian for MZMs located at the ends of the nanowires. In this framework, electron tunneling from the quantum dots into the wires occurs via MZMs ${\gamma_1}, {\gamma_2}, {\gamma_3}, {\gamma_4}$, as shown in \Cref{fig:tetron}(a). Within this picture, the observed bimodality in the quantum capacitance signal can be attributed to stochastic switching of fermion parities associated with the operators $X\equiv i\gamma_1 \gamma_3$ and $Z\equiv i\gamma_1 \gamma_2$ (or, equivalently,
$Z\equiv \pm i\gamma_3 \gamma_4$
for even/odd total parity).

Our minimal effective low-energy model takes the form:
\begin{align}
    H = & \; i\sum_{i<j} E_{ij}\gamma_i \gamma_j 
    + \sum_{i=1,L} \left[ \varepsilon_{\QD{i}} + \delta {\varepsilon_i} (t) \right] d^\dagger_i d_i \nonumber \\
    & + \sum_i \bigl[({t_\X^{(1)}}\,\gamma_1 + {t_\X^{(3)}} e^{i\varphi_\X}\, \gamma_3) d_1
    + \text{H.c.}\bigr] \nonumber \\
    & + \sum_i \bigl[ ({t_\Z^{(3)}}\, \gamma_3 + {t_\Z^{(4)}}e^{i\varphi_Z}\, \gamma_4) d_L
    + \text{H.c.} \bigr],
\end{align}
where $E_{ij}$ describes the splitting energy of a given pair of MZMs; $\varepsilon_{\QD{i}}$ denotes the respective detuning energies of the dots;
$\delta {\varepsilon_i}(t)$ represents
fluctuations in the detuning energy
of dot $i$;
$t_\X^{(1,3)}$, $t_\Z^{(3,4)}$ are the effective MZM-dot couplings; and $\varphi_{\X,\Z}$ are phases tuned via the fluxes through the
$X$, $Z$ loops. In \Cref{app:model}, we describe how this
two-dot model follows from a four-dot model. In either measurement configuration, noise --- such as charge noise leading to fluctuations of the QD detuning and Majorana splitting~\cite{Knapp18a} --- plays an important role as it can lead to transitions that are otherwise absent at zero temperature. Here, we focus on the impact of charge noise, which we model through fluctuations of quantum dot detuning $\delta {\varepsilon_i}(t)$~\cite{Boutin25}. Detuning is one of the largest scales and this noise mechanism can be expected to be one of the dominant mechanisms; other noise mechanisms will lead to qualitatively similar behavior.

\begin{figure}
    \centering
    \includegraphics{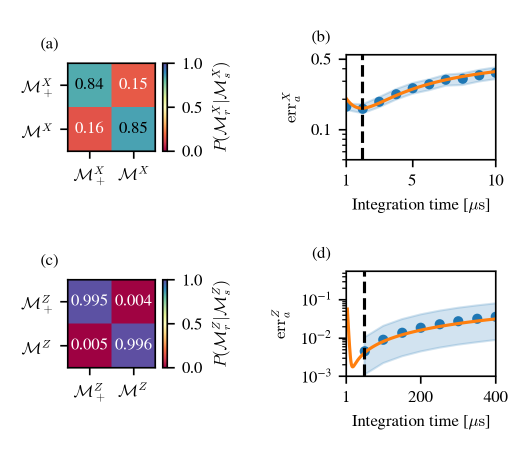}
    \caption{(a) The conditional probabilities for the outcome of an $X$ loop measurement given the outcome of an immediately preceding measurement using the time record shown in \Cref{fig:xmpr}. Here, $\mathcal{M}_r^P$ denote measurement outcomes; see \Cref{app:error_model} for details. Using the same data, (b) shows the probability ${\sf{err}_a}$ that two successive $X$ measurements give different results, plotted as a function of integration time (which can be adjusted by coarsening of the time record, i.e. averaging over some number of adjacent points). The optimal integration time of $\SI{2}{\micro\second}$ used for panel (a) is marked by a vertical dashed line. The orange line is a comparison to a noise model [\Cref{eqn:erra-expression}] using the lifetime $\tauX$ and SNR of 1.35 in $\SI{1}{\micro\second}$ as extracted from \Cref{fig:xmpr}(e) \cite{SNR_footnote}. The blue shading of ${\sf{err}_a}$ marks the $95\%$ confidence interval based on the length of the time record. Panels (c) and (d) plot the same quantities as in (a),(b) but for the $Z$ loop measurement. The orange line in (d) is again a comparison to a noise model using the lifetime $\tauZ$ and SNR of 0.85 in $\SI{1}{\micro\second}$ as extracted from a GMM fit to \Cref{fig:zmpr}(d). The optimal integration time is the timestep resolution of the data, which is $\SI{50}{\micro\second}$.}
    \label{fig:error_metrics}
\end{figure}

In this model, the observed RTS in $X$ or $Z$ basis measurements arise from fluctuations in fermion parity within the corresponding loops. These parity changes can result from quantum or thermally-activated fermion tunneling between MZM pairs within the tetron, or from extrinsic tunneling events to/from the ground (e.g., quasiparticle poisoning)~\cite{Knapp18a, Aghaee25}.
Indeed, the absorption of an above-gap excited quasiparticle by the MZMs can induce a fermion parity flip in the $X$, $Z$, or both loops, depending on the tunneling pathway.
There is a strong asymmetry between the hybrid wires' topological coherence length and gap, and the backbone's Al coherence length and gap.
In our model, this asymmetry leads to strong asymmetry between $\tauX$ and $\tauZ$, as tunneling across the backbone is negligible and, consequently, the leading contribution to $1/\tauZ$ is poisoning by external quasiparticles~\cite{Aghaee25}.
Conversely, the dominant contribution to
$1/\tauX$ appears to be inter-MZM fermion tunneling rather than a thermally-activated process,
as evidenced by the pronounced dependence on wire plunger voltages observed in \Cref{fig:xmpr}(i). This is because MZM couplings are expected to be highly sensitive to wire plunger tuning, whereas the topological gap is only weakly dependent on it.

Focusing on the effect of intra-wire MZM coupling, we estimate that the parity flip rate in the $X$ measurement configuration~--- assuming $E_{12}\approx E_{34}$~--- is given by ($\hbar=1$):
\begin{equation}
    \!\!\! \frac{1}{\tauXud} \approx
    S_\mathrm{N} (\pm \Delta E) \frac{E_\M^2}{2 t^2} \!
    \biggl[\Theta(M) \!+ \!\frac{t^2 \delta \phi^2 \! + \! 4x^2\!}{2 \Ed^2} \Theta(-M)\biggr] 
    \label{eq:xrate}
\end{equation}
in the limit $|\delta \phi| \ll |x|/t \ll |\Ed|/t \ll 1$; see \Cref{app:model} for derivations and expressions in other limits.
Here, $\Ed\equiv\varepsilon_{\mathrm{QD}_1}>0$, and $\Theta(M)$ is the Heaviside function, where $M$ denotes the total fermion parity of the dot-tetron system. The parameters are defined as follows: 
the average intra-wire MZM coupling is $E_\M = (E_{12} + E_{34}) / 2$; 
the average QD-MZM tunnel coupling is $t = ({t_\X^{(1)}} + {t_\X^{(3)}})/2$; 
the tunnel coupling asymmetry is $x = ({t_\X^{(1)}} - {t_\X^{(3)}}) / 2$;
the deviation from optimal flux bias is $\delta\phi = \varphi - \pi/2$;
the spectral function of fluctuations in the detuning of the dot is $S_\mathrm{N}(\omega)$ \cite{Aghaee25}; and
the energy difference between the two lowest energy eigenstates is $\Delta E$, see \Cref{app:model}.

Deep in the topological phase, the switching time $\tau_X$ is expected to grow exponentially with the topological gap $\Delta_T$ and the topological superconducting coherence length $\xi_T$, since the Majorana splitting scales as $E_M \propto \Delta_T A(k_F L) \, \exp\!\left(-\frac{L}{\xi_T}\right)$, where $L$ is the wire length, $k_F$ is the Fermi velocity and $A(k_F L)$ is a dimensionless function of $k_F L$~\cite{Meng09,DasSarma12}. In the clean limit, $\xi_T = v_F/\Delta_T$, where $v_F$ is the Fermi velocity, while in the presence of disorder, the coherence length $\xi$ can be strongly renormalized~\cite{Motrunich01,Brouwer11b}. Achieving fault-tolerant quantum computation requires $\frac{L}{\xi} \sim 10$~\cite{Aasen25}. In our current devices, $\frac{L}{\xi} \gtrsim 1$, so both the prefactor and the exponent are relevant. The prefactor $A(k_F L)$ introduces device-to-device variability, making the splitting energy sensitive to mesoscopic fluctuations and gate tuning.

From \Cref{eq:xrate}, we observe that when $E_{12}$ and $E_{34}$ are both finite and comparable, there is a cancellation effect that enhances $\tauX$ in one global parity sector --- particularly when the system is tuned to small $\delta\phi$, $x$. 
Away from optimal tuning we observe $\tauX \sim \SIrange{2}{5}{\micro\second}$ which is consistent with $E_\M \sim \SIrange{0.1}{0.3}{\micro\electronvolt}$ using the model presented here, see \Cref{app:qpp_x} for details. Additional qualitative features in \Cref{fig:xmpr} support this interpretation: fluctuations in global parity between time records can lead to nearby flux points with significantly different histograms, (see the rapidly-varying part of
the excess kurtosis in
\Cref{fig:xmpr}(c)), and diagonal features in the wire plunger-wire plunger scan (panel (i)) could indicate the condition of $E_{12}(V_\mathrm{WP1})\sim E_{34}(V_\mathrm{WP2})$. Moreover, as explained in \Cref{app:model}, a non-zero $E_\M$ can also induce a finite $\dCq$ as a function of flux which, together with an $ E_\M$-induced sensitivity to quasiparticle poisoning (QPP), can lead to the increased width apparent in the histogram of panel (g) relative to panel (e). We discuss these effects in more detail in \Cref{app:model} and \Cref{app:qpp_x}.

A key metric in quantifying the performance of measurement-based qubits is the operational assignment error $\erra$~\cite{Aasen25} (see also \Cref{app:error_model}), which quantifies the probability of
failing to obtain the same result
when repeating the same type of measurement. \Cref{fig:error_metrics} shows the error metrics of the $X$ and $Z$ loop measurements when interpreting each data point as an independent measurement. We find errors $\errXa = 16\%$, $\errZa = 0.5\%$  corresponding to the time records shown in \Cref{fig:xmpr,fig:zmpr} when applying optimal integration times of $\SI{2}{\micro\second}$ and $\SI{50}{\micro\second}$, respectively \footnote{Extrapolating our error model [orange line in \Cref{fig:error_metrics}(d)] to shorter integration times we expect $\errZa \approx 0.2\%$ at optimal integration times of $\sim \SI{20}{\micro\second}$.}. The asymmetry in errors is due to the difference in lifetimes $\tauZ \gg \tauX$. Comparing the data to an error model based on finite SNR (extracted from the histograms of the time records) and finite lifetimes (extracted from the autocorrelation function or dwell time distribution) we find good agreement with the experimental data without further fitting parameters, see \Cref{app:error_model} for more details.

\paragraph{Discussion and outlook.} \label{sec:alternatives}

The results presented here represent a significant advance toward the increased device and system complexity required for the practical operation of topological qubits. We have characterized the operational assigment errors for $X$ and $Z$ measurements and extracted corresponding timescales for parity switching. Notably, we find that $\tauX$ and $\tauZ$ differ significantly, consistent with a model in which MZMs have small but finite splitting energies. Thus far, we have explored only a limited subset of the parameter space identified by the TGP--—for example, using only a single in-plane magnetic field value. Nevertheless, extrapolating from the wire plunger scan in \Cref{fig:xmpr}(i), we find that the pronounced variation in $\tauX$ is consistent with mesoscopic fluctuations, which are expected to dominate the MZM splitting in current devices~\cite{Brouwer11b}. Looking ahead, improvements in materials and fabrication techniques can yield larger topological gaps and shorter coherence lengths. These enhancements would exponentially suppress the typical MZM splittings and expand the usable phase space~\cite{Adagideli14, Aghaee23}. 

As in any interferometric measurement of low-energy states~\cite{Aghaee25}, one cannot 
 rule out non-topological explanations: trivial models which may include additional ``hidden'' MZMs 
~\cite{Janvier15, Hays18, Hays21, Wesdorp23, Elfeky23a,Kells12, Liu17, Vuik19, Valentini21} 
can be fine-tuned to give similar results~\cite{Aghaee25}.
In \Cref{app:alternative_scenarios},
we analyze several non-topological
models in detail and discuss the
degree of fine-tuning that is necessary for a non-topological model
to be consistent with our data.
As continued progress is made on the roadmap towards a topological quantum computer~\cite{Aasen25}, results
on more complex devices with smaller
$\errXa$, $\errZa$ can continue to render non-topological explanations less and less tenable.

The next key step in this roadmap
is rapid sequences of
$X$ and $Z$ measurements, which can
show that these two measurements are not
only distinct but, in fact,
do not commute. With two tetrons,
measurements of the total parity
of four MZMs are possible, enabling entanglement
and braiding.

\section*{Data and code availability}
The datasets associated with the figures in this paper are available on Zenodo (\url{https://doi.org/10.5281/zenodo.16987493}). The source code that performs the analysis and generates the figures is available on GitHub (\url{https://github.com/microsoft/microsoft-quantum-tetron-lifetimes}).

\begin{acknowledgments}
We thank Jason Alicea, Leonid Glazman,
Sankar Das Sarma, and Jaydeep Sau for discussions.
We are grateful for the contributions
of David Pinon and Ramin Sadaghiani at an
early stage in this project.
We thank Julian Herring for assistance with manuscript preparation and Edward Lee, Todd Ingalls and Connor Gilbert for assistance with the figures.
\end{acknowledgments}

\appendix
\crefalias{section}{appsec}

\renewcommand{\thefigure}{A\arabic{figure}}
\setcounter{figure}{0}

\section{Theoretical model}
\label{app:model}

\begin{figure}[b]
\centering
\includegraphics[width=\columnwidth]{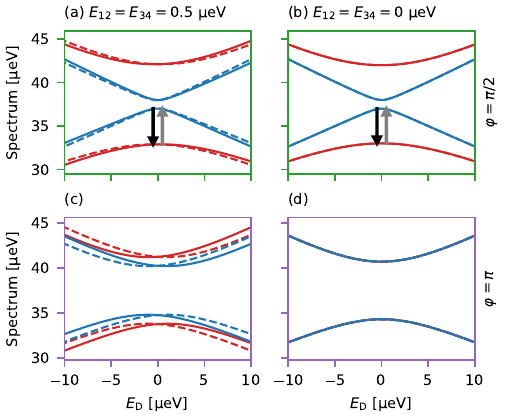}
\caption{
Energy spectrum of the low-energy effective model as a function of detuning energy. The top row [panels (a,b)] shows the spectrum at the flux corresponding to maximal visibility (dashed green line in \Cref{fig:spectrum_flux}), while the bottom row [panels (c,d)] shows the spectrum for flux corresponding to minimal visibility (dashed purple line in \Cref{fig:spectrum_flux}). The left panels (a,c) show finite $E_{12} = E_{34} = \SI{0.5}{\ueV}$, while the right panels (b,d) corresponds to $E_{12}, E_{34} = 0$.
The black and gray arrows indicate noise-induced transitions between parity eigenstates as evaluated from \Cref{eq:fermi_golden_rule} for $\varphi\sim \pi/2$. Solid (dashed) curves correspond to the even (odd) global parity sector $M=\pm1$.
}
\label{fig:spectrum_detuning}
\end{figure}

\begin{figure}
\centering
\includegraphics[width=\columnwidth]{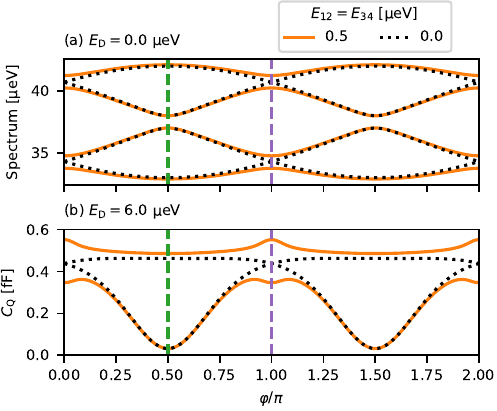}
\caption{
(a)~Spectrum of Hamiltonian \Cref{eqn:eff-Hamiltonian1} as a function of flux for $E_{ij}=0$ (dotted black curves) and for $E_{12} = E_{34} = \SI{0.5}{\ueV}$ (solid orange curves), corresponding to vanishing and non-zero Majorana overlap, respectively.
Parameters are chosen to be at charge degeneracy ($\Ed = 0$) and in the nearly balanced interferometer regime ($t_\X^{(1)} = \SI{2}{\ueV}$, $t_\X^{(3)} = \SI{2.5}{\ueV}$).
(b)~Quantum capacitance in the static (small readout tone frequency and amplitude) and low-temperature limit. Both panels use fixed even total parity $M=-1$.
}
\label{fig:spectrum_flux}
\end{figure}

We now introduce a theoretical model for the quantum capacitance measurements performed here, and the jump processes that can occur in the measurement configuration. 
To this end, we introduce an effective low-energy model for the wire and adjacent quantum dots and derive its spectrum and eigenstates. The quantum capacitance measurement can be viewed as a quantum non-demolition measurement that projects onto eigenstates of this Hamiltonian; see the discussion, e.g., in Refs.~\onlinecite{Karzig17, Aghaee25, Boutin25, Sau24} for further details on this. 
We then argue that noise on the quantum dots can lead to transitions between eigenstates of the system, as illustrated by the arrows in the top panels of \Cref{fig:spectrum_detuning}. 
These transitions can be viewed as akin to relaxation/excitation processes between computational basis states in conventional qubits; however, it is important to emphasize that here, they occur specifically in the measurement configuration, whereas the computational states in a tetron are near-degenerate when the qubit is isolated from the adjacent quantum dots; see Ref.~\onlinecite{Aasen25} for further discussion of this point.

We begin with a model for four MZMs $\gamma_1$, $\gamma_2$, $\gamma_3$, $\gamma_4$ and four dots, 1, 3, 4, and L. We ignore dot 2, which is decoupled from the wire and the other dots during the measurements reported in this paper.
The Hamiltonian for this system takes the form:
\begin{align}
    H = & \; i\sum_{i<j} E_{ij}\gamma_i \gamma_j + \sum_i \varepsilon_{\QD{i}} d^\dagger_i d_i \nonumber\\ & + \sum_i \tm{i} \gamma_i d_i + \sum_{[i,j]_d} t_{ij}d^\dagger_i d_j + \text{H.c.},
    \label{eqn:eff-Hamiltonian}
\end{align}
where $E_{ij}$ describes the splitting energy of a given pair of MZMs; $\varepsilon_{\QD{i}}$ denotes the respective detuning energies of the 4 dots; $\tm{i}$ is the coupling between the MZMs and their adjacent dots; and $t_{ij}$ is the dot-dot coupling between neighboring dots (denoted by $[i,j]_d$). The $X$ measurement configuration corresponds to a measurement loop involving $\gamma_1$,$\gamma_3$, dot 1, and dot 3 while the $Z$ measurement configuration involves $\gamma_3$,$\gamma_4$ and dots 3, L, and 4. 
While $t_\X^{(1)} = \tm{1}$, the other effective couplings are mediated through detuned dots 3 and 4: $t_\X^{(3)} \sim t_{13} \tm{3}/\Delta_{3}$, $t_\Z^{(3)} \sim t_{3L} \tm{3}/\Delta_{3}$, $t_\Z^{(4)} \sim t_{4L} \tm{4}/\Delta_{4}$, where $\Delta_{3,4}$ are the detuning energies of dots 3 and 4 relative to charge degeneracy point.
We describe how these couplings are tuned in \Cref{app:device_tuning}.
The loops are connected or disconnected by changing the tunnel couplings $t_{13}$ and $t_{3L}$ between the tunneling and the fully closed regime, respectively.

We now focus on the case in which $t_{3L} = t_{4L} = 0$ and, hence, ${t_\Z^{(3)}} = {t_\Z^{(4)}} = 0$. 
Since dot L is disconnected from the bottom wire, we can ignore it and introduce an effective low-energy model for the tetron system comprising the four MZMs and dot 1, which has occupancy ${d^\dagger}d$ and detuning energy $\Ed \equiv \varepsilon_{\QD{1}}$. 
This simplified framework enables an analytical evaluation of qubit error rates in the $X$ measurement configuration and reveals how these rates depend on MZM coupling, magnetic flux, and dot detuning. 
The effective Hamiltonian reads 
\begin{align}
\label{eqn:eff-Hamiltonian1}
    H =& \, i\sum_{i<j} E_{ij}\gamma_i \gamma_j + \Ed \left(d^\dagger d-\frac{1}{2}\right) \\ & + {t_\X^{(1)}} \gamma_1 d +
    {t_\X^{(3)}}e^{i\varphi} \gamma_3 d  + \text{H.c.}\,, \nonumber\\
    H_\mathrm{noise} =& \,\delta \varepsilon_1 (t) d^\dag d.
\end{align}
Here, $E_{ij}$ denotes the MZM coupling energies, and $\Ed$ represents the dot detuning, assuming a single non-degenerate level on the dot (i.e., large level spacing).
We set $t_\X^{(1)}$, $t_\X^{(3)}$ to be real and capture the phase dependence of the tunneling elements via $\varphi = 2\pi \Phi/(h/e)+\varphi_0$ in terms of the applied flux $\Phi$ and an flux-independent offset $\varphi_0$. 
We also assume that the topological gaps in both wires are large compared to the temperature. 
Charge noise acting on the dot is modeled by the stochastic variable $\delta \varepsilon_1(t)$, characterized by the spectral function $S_\mathrm{N}(\omega) = \gamma_\mathrm{eff} / [1 + \exp(-\omega / \kB T)]$ with $\gamma_\mathrm{eff}$ being effective noise strength~\cite{Boutin25}. Henceforth, we neglect $E_{13}$, $E_{14}$, $E_{23}$, and $E_{24}$ assuming that the trivial superconducting backbone
has a short coherence length and suppresses fermion tunneling processes that cross the backbone. 

We now derive the effective Hamiltonian for a fixed parity sector of the combined dot-tetron system. We adopt the basis convention $\ket{k} \equiv \ket{\text{QD}} \otimes \ket{i\gamma_1 \gamma_3} \otimes \ket{i\gamma_2 \gamma_4}$, where the state is defined over the quantum dot and two fermion parity sectors of the MZMs. In this basis, the effective low-energy Hamiltonian takes the form:
\begin{widetext}
\begin{align}
H_\M = \left ( 
\begin{array}{cccc}
     \Ed / 2 & {t_\X^{(1)}}-i {t_\X^{(3)}}e^{-i \varphi} & -i (E_{12}- E_{34} M) & 0 \\
     {t_\X^{(1)}}+i {t_\X^{(3)}}e^{i \varphi} & - \Ed / 2 & 0 & i (E_{12}+ E_{34}M)  \\
     i (E_{12}-E_{34} M)  & 0 &  \Ed / 2   &  {t_\X^{(1)}}+i {t_\X^{(3)}}e^{-i \varphi} \\
     0 &  -i (E_{12}+E_{34}M) &  {t_\X^{(1)}}-i {t_\X^{(3)}}e^{i \varphi} & - \Ed / 2  \\
\end{array} \right).
\label{eq:HM}
\end{align}
\end{widetext}
Here, $M = \pm 1$ denotes the total fermion parity of the dot-tetron system. 
For the even parity sector ($M = -1$), the basis states are $\ket{1,0,1}$, $\ket{0,1,1}$, $\ket{1,1,0}$, and $\ket{0,0,0}$; for the odd parity sector ($M = +1$), the basis states are $\ket{1,0,0}$, $\ket{0,1,0}$, $\ket{1,1,1}$, and $\ket{0,0,1}$. 
The upper-left and lower-right blocks of the Hamiltonian correspond to different parity sectors of the dot–MZM$_{13}$ loop. 
The energy spectrum of Hamiltonian \eqref{eq:HM} as a function of magnetic flux and dot detuning is shown in \Cref{fig:spectrum_detuning} and \Cref{fig:spectrum_flux}. 
In the absence of MZM splittings, the lowest energy states are given by $E_\X = -\frac{1}{2}\sqrt{\Ed^2 + 4 ({t_\X^{(1)}}^2+{t_\X^{(3)}}^2+2X{t_\X^{(1)}}{t_\X^{(3)}}\sin \varphi )}$, where $X$ denotes the eigenvalues of the fermion parity operator $i\gamma_1\gamma_3$. 
In this idealized regime, the even and odd total parity sectors $M = \pm 1$ are degenerate. 
However, finite MZM splittings lift this degeneracy.  
Additionally, nonzero $E_{12/34}$ induce avoided level crossings at $\varphi=\pi$, resulting in a finite quantum capacitance shift $\dCq$ at that point, see \Cref{fig:spectrum_flux}.     

We now analyze the effect of finite MZM splittings on noise-induced transitions between the lowest energy states. 
Assuming $E_{12}, E_{34} \ll {t_\X^{(1)}}, {t_\X^{(3)}}$ and $|\varphi \pm \pi/2| \ll 1$, the Hamiltonian $H_\M$ can be block-diagonalized using a Schrieffer–Wolff transformation. 
In the resulting ``dressed'' state basis, the effective Hamiltonians take the form $\tilde{H}_\M = H_\M^0 + V$, where $H_\M^0$ corresponds to the $E_{12,34} = 0$ limit, and the noise Hamiltonian is given by $\tilde{H}_\mathrm{N}=\delta \varepsilon_1 (t) \hat{K}$. 
The perturbation and the noise operator are expressed as:
\begin{widetext}
\begin{align}
&V=\left(\!
\begin{array}{cccc}
 -A_{11}(M) \Ed & i E_{12} A_{12}- E_{34} A_{34} & 0 & 0\\
 -i E_{12} A^*_{12}- E_{34} A_{34}  & A_{11}(-M) \Ed & 0 & 0 \\
 0 & 0 & A_{11}(M) \Ed & i E_{12}A_{12}+ E_{34} A_{34} \\
0 & 0 & -i E_{12} A^*_{12}+ E_{34} A_{34}  & - A_{11}(-M) \Ed\\
\end{array}
\!\right),\\
&\nonumber\\
\!&\!\hat{K} = \left(\!
\begin{array}{cccc}
    1/2 & 0 & 0 & A_{12}+ i M A_{34}\\
    0 & -1/2 & -A_{12}^*-i M  A_{34} & 0 \\
    0 & -A_{12}+i M A_{34} & 1/2 & 0 \\
    A_{12}^*-i M A_{34} & 0 & 0 & -1/2 \\
\end{array}
\!\right).
\end{align}
\end{widetext}
Here the coefficients are defined as: $A_{11}(M) = (E_{12} - E_{34}M)^2 \csc\varphi / 4 t_\X^{(1)} t_\X^{(3)}$, $A_{12} = E_{12} (\cot\varphi -i) / 2 t_\X^{(1)}$, and $A_{34} = E_{34} \csc\varphi / 2 {t_\X^{(3)}}$. 
Finite MZM splittings introduce two key effects: 
(a)~the degeneracy between the $M=\pm 1$ global parity sectors is now lifted by an energy $\approx \Ed E_{12} E_{34} \csc\varphi / t_\X^{(1)} t_\X^{(3)}$, which may lead to a measurable difference in quantum capacitance between the sectors. 
Notably, the effective Hamiltonian $\tilde{H}_{M}$ is symmetric under the transformation $X \rightarrow - X$ and $\varphi \rightarrow \varphi +\pi$;  
(b)~within each fixed parity sector $M$, the noise Hamiltonian $\tilde{H}_\mathrm{noise}$ contains off-diagonal terms, enabling intra-sector transitions between the lowest-energy states of $\tilde{H}_\mathrm{noise}$, denoted
$\ket{\pm}$. 

To leading order in the MZM splitting, the transition rate can be computed using Fermi's Golden Rule: 
\begin{equation}
    \frac{1}{\tauXud} = 
    S_\mathrm{N}\left(\pm \Delta E \right) \bigl| \bra{+}\hat{K}\ket{-}\bigr|^2.
    \label{eq:fermi_golden_rule}
\end{equation}
where $\Delta E = E_+ - E_-$ while $E_\pm$ and $\ket{\pm}$ denote the eigenvalues and eigenstates of the lowest-energy doublet, which correspond to different eigenvalues of the operator $X$ in the limit that the
dot is decoupled. 
For the sake of simplicity, we focus on the regime $|\varphi - \pi/2|\ll 1$, and parametrize $E_{12,34} = E_\M\mp y$, assuming $E_\M \gg |y|$. 
Furthermore, we neglect thermally-activated processes assuming temperature is much smaller than topological gaps in upper and lower wires and consider transitions that require the tunneling of a fermion between the left and right pairs of MZMs. 

Since the off-diagonal elements of $\hat{K}$ scale linearly with $E_\M$, the resulting transition rates $1/\tauXud$ are proportional to $E_\M^2$ in the leading order. In the regime of small flux detuning from the optimal point and a well-balanced interferometer, this yields the following expression for the transition rate:
\begin{widetext}
\begin{align}
\centering
\frac{1}{\tauXud} \approx S_\mathrm{N} \left(\pm \Delta E \right)\frac{E_\M^2}{4 t^2} \left \{ \begin{array}{ll}
  1 +M \cfrac{\Ed}{|x|}  , \, & \, |\delta \phi|  \ll |\Ed|/t \ll |x|/t \ll 1, \vspace{1.5mm} \\ 
  2 \Theta(M) + \cfrac{2(t^2 \delta \phi^2+4x^2)}{\Ed^2} \,\Theta(-M) ,   & |\delta \phi| \ll |x|/t\ll |\Ed|/t \ll 1,  \vspace{1.5mm} \\ 
  \cfrac{16 t^2}{\Ed^2}\,\Theta(M) + \cfrac{64 t^4 (4x^2+t^2\delta \phi^2) }{\Ed^6} \, \Theta(-M) ,   & |\delta \phi| \ll  |x|/t\ll  1 \ll |\Ed|/t. \\ 
\end{array}\right. 
\label{eq:rate}
\end{align}
\end{widetext}
Here $\Ed > 0$, $t = ({t_\X^{(1)}} + {t_\X^{(3)}})/2$ is the average tunnel coupling, $x = ({t_\X^{(1)}} - {t_\X^{(3)}})/2$ is the tunnel coupling asymmetry, $\delta\phi = \varphi - \pi/2$ is the deviation from the optimal flux bias.
The result for $\Ed < 0$ can be obtained by applying the transformation $M \rightarrow -M$ in Eq.~\eqref{eq:rate}. This symmetry arises from the fact that the off-diagonal matrix elements in Eq.~\eqref{eq:HM}  exhibit constructive or destructive interference depending on the sign of the total parity $M$. Finally, in the opposite limit where $|y| \gg E_\M$, the corresponding rate can be obtained by substituting $E_\M \rightarrow y$, $M\rightarrow -M$ in Eq.~\eqref{eq:rate}. This reflects the symmetry of the Hamiltonian \eqref{eq:HM} under parity inversion.

To analyze the quantum dynamics at fixed total parity $M$, we model transitions 
between the two lowest-energy states $\ket{\pm}$ using the rate equations
\begin{align}
\frac{d P^{(M)}_{-}}{dt} &= \frac{1}{\tau_{X\downarrow}}\, P^{(M)}_{+}
                          - \frac{1}{\tau_{X\uparrow}}\, P^{(M)}_{-},\\
\frac{d P^{(M)}_{+}}{dt} &= \frac{1}{\tau_{X\uparrow}}\, P^{(M)}_{-}
                          - \frac{1}{\tau_{X\downarrow}}\, P^{(M)}_{+},
\end{align}
with $P^{(M)}_{\pm}$ the probabilities to occupy $\ket{\pm}$. Probability conservation
implies $P^{(M)}_{-}+P^{(M)}_{+}=1$.

Linearizing about the steady state, we introduce 
$P^{(M)}_{-}(t)=\bar P^{(M)}_{-}+\delta P^{(M)}_{-}(t)$ and 
$P^{(M)}_{+}(t)=\bar P^{(M)}_{+}+\delta P^{(M)}_{+}(t)$, where 
$\bar P^{(M)}_{\pm}$ are the stationary populations and 
$\delta P^{(M)}_{+}(t)=-\delta P^{(M)}_{-}(t)$. The fluctuations obey
\begin{align}
\frac{d\,\delta P^{(M)}_{-}}{dt}
= -\Gamma\,\delta P^{(M)}_{-} + \xi(t),
\qquad 
\Gamma \equiv \tau_{X\uparrow}^{-1}+\tau_{X\downarrow}^{-1},
\end{align}
driven by a Langevin source $\xi(t)$ with white-noise correlations
$\langle \xi(t)\,\xi(t')\rangle = A_0\,\delta(t-t')$. For a two-state system, the equilibrium distribution is binomial, so the variance of fluctuations is $\langle [\delta P^{(M)}_{-}]^2\rangle = \bar P^{(M)}_{-}\bar P^{(M)}_{+}$. Within the white-noise approximation, matching the equal-time variance to this equilibrium result fixes the noise strength as $A_0 = 2\,\Gamma\,\bar P^{(M)}_{-}\bar P^{(M)}_{+}$.

For an observable $\Cq$ that jumps by $\dCq^{(M)}$ between the two states, the (random-telegraph) autocorrelation of the fluctuation $\dCq(t)\propto \delta P^{(M)}_{-}(t)$ is
\begin{align}
    \mathcal{A}[\Cq]^{(M)}_\RTS(t-t')
    = \big(\dCq^{(M)}\big)^2\,\bar P^{(M)}_{-}\bar P^{(M)}_{+}\;
    e^{-\Gamma\,|t-t'|}.
\end{align}
Note that we defined $\tauX^{-1}$ as the average of the up and down rates and thus $\tauX \equiv 2/\Gamma$.
In the limit where the doublet is nearly degenerate $\bar P^{(M)}_{-} \approx \bar P^{(M)}_{+}=1/2$ 
(i.e., $\tau_{X\uparrow}\approx\tau_{X\downarrow}$), the prefactor simplifies to 
$\bar P^{(M)}_{-}\bar P^{(M)}_{+} = 1/4$, yielding
\begin{align}
\!\!\mathcal{A}[\Cq]^{(M)}_\RTS(t-t')
\!=\!\left(\frac{\dCq^{(M)}}{2}\right)^{\!2}
  \exp\!\left(-\frac{2|t-t'|}{\tauX}\right),
\end{align}
which was used in Eq.\eqref{eqn:autocorr} of the main text.

On time scales much longer than $\tauX$ the jumps $\Delta \Cq ^{(M)}$ average out to $\bar{\Cq}^{(M)}$. As shown above, finite Majorana couplings can lead to a total parity dependence of the average such that $\Delta C_{\Delta M}=\bar{\Cq}^{(+1)}-\bar{\Cq}^{(-1)}$ remains finite and it becomes possible to observe a second RTS with a slow time scale $\tau_\mathrm{s}\gg \tau_X$ given by total parity flips due to QPP.

This model is consistent with experimental observations and captures the dominant contribution to the $X$-measurement lifetime arising from charge noise and finite MZM splittings. The derived autocorrelation function and noise power provide a direct link between transition rates, MZM hybridization, and measurable time-domain signatures, enabling predictive estimates of the device performance.

\section{Additional data and comparison to theoretical model} \label{app:qpp_x}

\begin{figure}
\centering
\includegraphics[width=\columnwidth]{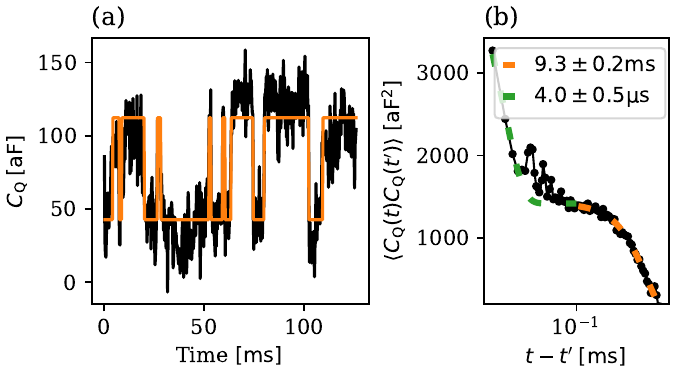}
\caption{
Analysis of longer time records showcasing the existence of two timescales. (a)~The $\SI{120}{\milli\second}$ long coarsened time record (black) has discrete RTS transitions with a timescale much longer than $\SI{10}{\micro\second}$ as evident by the GMM classification (orange) when coarsening to an effective integration time of $\SI{250}{\micro\second}$. 
(b)~Autocorrelation of the time record of (a) without any applyied coarsening (integration time $\SI{1}{\micro\second}$). 
One can observe two regimes of applicability for the RTS autocorrelation model \eqref{eq:autocorrelation}: a $\SI{4.0 \pm 0.5}{\micro\second}$ timescale (green) of similar order to the timescales in \Cref{fig:xmpr} and a $\SI{9.3 \pm 0.2}{\milli\second}$ timescale (orange) consistent with QPP and the data from figure \Cref{fig:zmpr}. 
The extracted signals are $\dCq = \SI{109 \pm 1}{\atto\farad}$ and $\Delta C_{\Delta M} = \SI{75 \pm 6}{\atto\farad}$, respectively.
}
\label{fig:long_time_record}
\end{figure}

\paragraph{Multiple timescales in long time records.}

\begin{figure}
\centering
\includegraphics[width=\columnwidth]{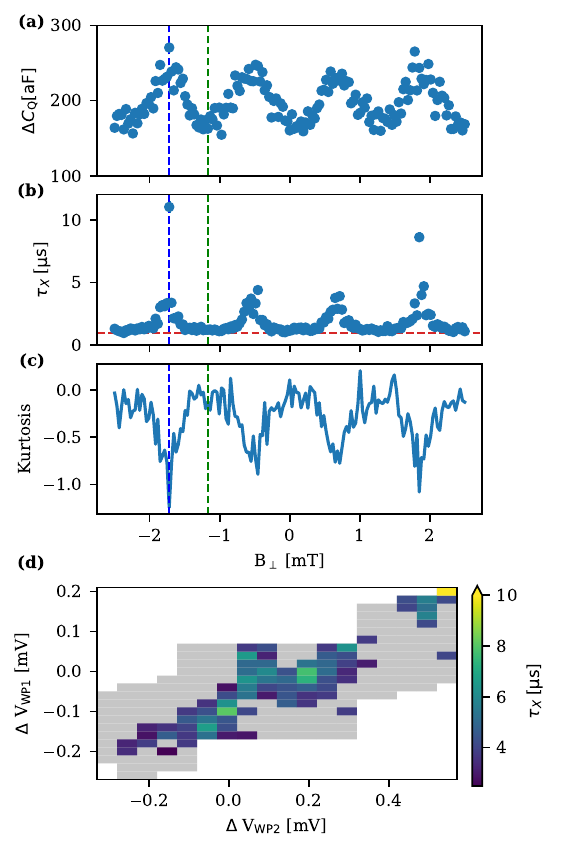}
\caption{
Autocorrelation analysis over a wider range in parameter space. 
(a,b)~Extracted parameters $\dCq$ and $\tauX$ from fits of the autocorrelation function similar to \Cref{fig:xmpr}(h) but over the full range of $\Bperp$. 
The dashed red line in (b) indicates the measurement integration time. 
For easier comparison to \Cref{fig:xmpr} panel~(c) replots the kurtosis for the corresponding time records. 
(d)~Dependence of $\tauX$ on wire plunger tuning over the same data and $V_{\mathrm{WP1}}$-$V_{\mathrm{WP2}}$ range as in \Cref{fig:xmpr}(i). To improve fit reliability, only fits with $\mathrm{SNR}>1.0$ are taken into account as determined by the size of the fitted $\Delta C_Q$. The plot then shows the top 1\% of timescales for all detunings and $B_\perp$ values meeting the SNR requirement. The gray background indicates points where data was taken but the SNR requirement could not be met.  
}
\label{fig:timescales}
\end{figure}

To further investigate the predicted interplay of multiple time scales --- state flips occurring on the order of a few $\SI{}{\micro\second}$ and QPP on the order of several $\SI{}{\milli\second}$ --- we recorded long time traces of $\SI{120}{\milli\second}$ length in the $X$ measurement configuration, near the tuning point used in \Cref{fig:xmpr}. 
\Cref{fig:long_time_record} shows an example time record with significant coarsening and the corresponding autocorrelation function of the time record without applied coarsening. 
The coarsened data reveals slow RTS with a characteristic time scale of approximately $\SI{10}{\milli\second}$, consistent with the time scales observed in \Cref{fig:zmpr}, which we attribute to QPP. 
Note that generally one would expect the poisoning time scale of total parity $M$ to be faster than the poisoning of the Z loop alone due to the increased cross section for capturing QPs. 
In addition, the autocorrelation function of the data reveals a faster time scale of the order of a few $\SI{}{\micro\second}$ consistent with the time scales observed in the data of \Cref{fig:xmpr}. 
For these long time records both time scales can be fitted at the same time when assuming a model of the autocorrelation function of the form
\begin{eqnarray}
    \mathcal{A}[\Cq](t-t')&=& \sigma_{\Cq}^2 \delta(t-t') +\biggl(\frac{\dCq}{2}\biggr)^2 e^{-2|t-t'|/\tau_\mathrm{f}} \nonumber \\
    &&+ \biggl(\frac{\Delta C_{\Delta M}}{2}\biggr)^2 e^{-2|t-t'|/\tau_\mathrm{s}}
    \label{eq:autocorrelation}
\end{eqnarray}
describing two independent RTS with well-separated fast and slow time scales, $\tau_\mathrm{f}$ and $\tau_\mathrm{s} \gg \tau_\mathrm{f}$ respectively. Note that within the model presented in \Cref{app:model} the fast time scale can depend on the total parity $M$. To avoid overfitting features at fast times we used a simplified model in Eq.~\eqref{eq:autocorrelation} assuming $\dCq=\dCq^{(+1)}=\dCq^{(-1)}$ and $\tau_\mathrm{f}=\tau_{X,M=+1}=\tau_{X,M=-1}$. We expect $\Delta C_{\Delta M}$ to be most visible at phases $\varphi=\pi$ where $\dCq$ is minimal as shown in \Cref{fig:spectrum_detuning}. 
To focus on the long time scale we thus used a flux point where the oscillation of the standard deviation of the time record shows a local minimum with respect to $\Bperp$.
While in this regime $\dCq$ can be similar to $\Delta C_{\Delta M}$ as shown in \Cref{fig:long_time_record} we expect that in the regime of high visibility ($\varphi \approx \pi/2$) $\dCq$ dominates over $\Delta C_{\Delta M}$.

In conclusion, the long time record data is consistent with the theoretical model introduced in \Cref{app:model} and the separation of time scales by 3 orders of magnitude justifies studying their effect separately as done in \Cref{fig:xmpr} and \Cref{fig:zmpr}.

\paragraph{Extracting $\tauX$ over a wider range in parameter space.}

To estimate the typical values of residual coupling $E_\M$, it is useful to extract the timescale $\tauX$ away from the optimal tuning points. 
To this end, we fit the autocorrelation function --- similar to \Cref{fig:xmpr}(h) but optimized to be more robust to shorter time scales --- across a range of $\Bperp$. We observe that typical timescales away from optimal tuning are in the range of $\SIrange{2}{5}{\micro\second}$. 
Using the value of sub-optimal $\tauX^{-1}\sim \gamma_\mathrm{eff}E_\M^2/(2t)^2$ with $t \sim \SI{2}{\micro\electronvolt}$ and $\gamma_\mathrm{eff} \sim \SIrange{0.1}{1}{\giga\hertz}$ \cite{Boutin25}, we estimate $E_\M \sim \SIrange{0.1}{0.3}{\micro\electronvolt}$. Additionally, we extract $\tauX$ for various gate voltage configurations $(V_\mathrm{WP1},V_\mathrm{WP2})$ based on the data presented in \Cref{fig:xmpr}(i). The results are summarized in \Cref{fig:timescales}.

\section{Alternative scenarios} \label{app:alternative_scenarios}
So far, we interpreted the data presented in \Cref{fig:xmpr,fig:zmpr} in terms of a model in which the two horizontal wire segments are in the topological phase and the backbone is in the topologically-trivial gapped superconducting phase. The low-energy physics of this scenario can be described by 4 MZMs, one located at each end of the ``H" structure, as in \Cref{eqn:eff-Hamiltonian}. Alternative scenarios can be generated by considering different locations of the MZMs or incorporating additional Majorana modes in the low-energy Hamiltonian. They are, however, significantly constrained by the entirety of the data, comprising:
\begin{enumerate}
\item Device tune up using TGP on both wire segments, which entails stable ZBPs at all 4 ends of the device and a bulk transport gap in both wires; \label{it:tgp}
\item RTS of $\tau_X = \SIrange{2}{15}{\micro\second}$ time scale for $X$ measurement, depending on flux; \label{it:Xtime}
\item RTS visibility ($\dCq$) consistent with $h/2e$ period for $X$  measurement; \label{it:Xflux}
\item RTS of  $\tau_Z\sim \SI{12}{\milli\second}$ time scale for $Z$ measurement; \label{it:Ztime}
\item RTS visibility ($\dCq$) consistent with $h/2e$ period for $Z$ measurement; \label{it:Zflux}
\item Slower time scale $\sim \SI{10}{\milli\second}$ RTS in $X$ measurement with smaller amplitude $\Delta C_{\Delta M}$. \label{it:2times}
\end{enumerate}

\begin{figure}
    \centering
    \includegraphics[width=0.9\linewidth]{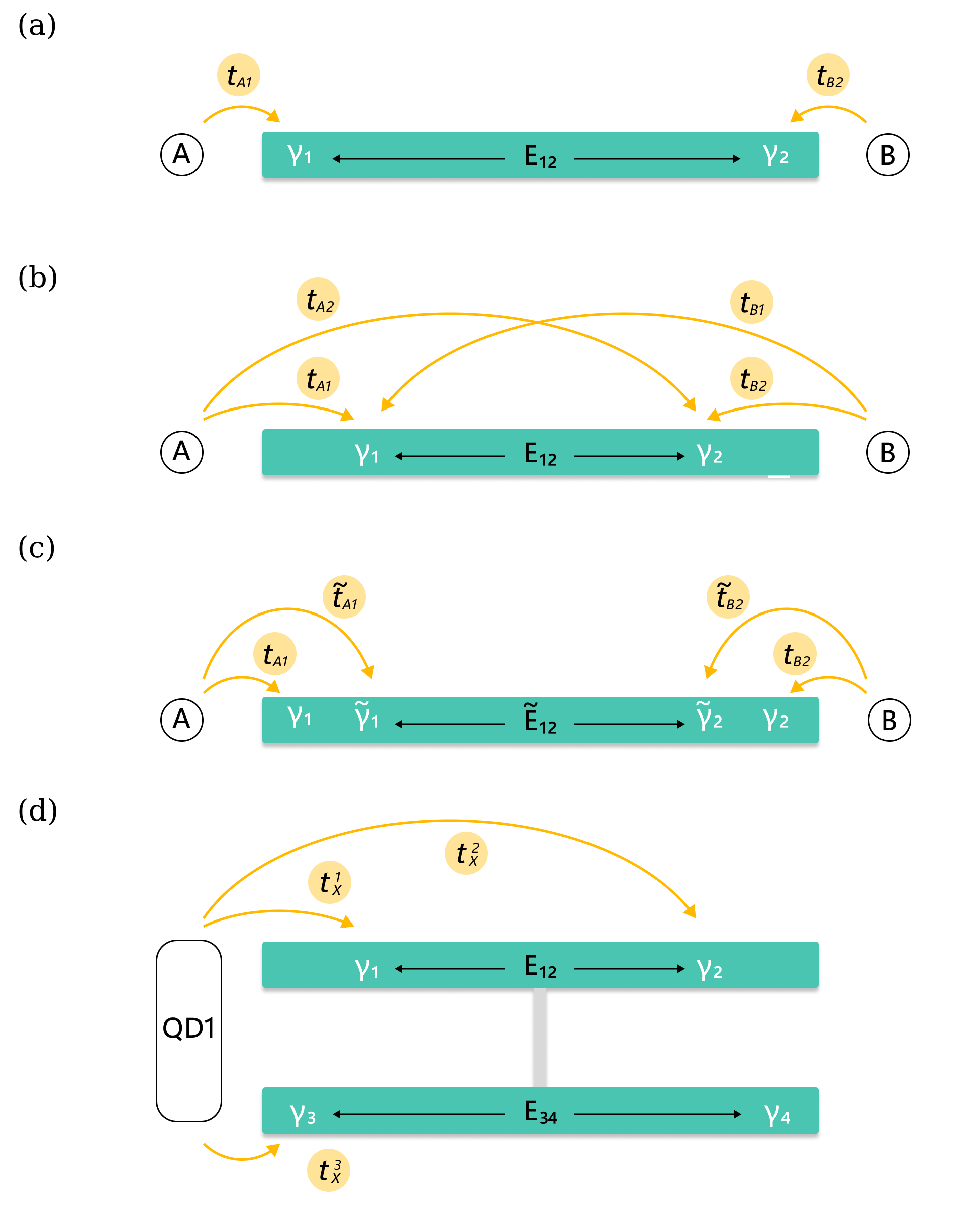}
    \caption{Overview of different scenarios for low energy wire states. (a) Depicts the topological scenario of a nanowire with Majorana modes localized at the ends of the wire with small residual coupling $E_{12}$. (b) shows the case of a general spin-resolved Andreev mode spread out over the wire. (c) is an example of a non-topological scenario with two quasi Majorana modes localized at each end of the wire. These single wire scenarios could appear in various combinations for the top and bottom wire of the tetron device. (d) shows an abstracted view of the X loop measurement configuration where the upper and lower wires are described by scenarios (b) and (a), respectively. The effect of the other dot, QD3, that involved in the measurement loop is captured by $t_\X^{(3)}$ as described in \Cref{app:model}.}
    \label{fig:alt_scen}
\end{figure}
Due to the presence of stable ZBPs at all four ends of the device, we limit the discussion to scenarios where low-energy modes can be accessed from both sides of both nanowires. For a given wire, we consider three scenarios, see \Cref{fig:alt_scen}: (a) the topological scenario, (b) a spin-resolved Andreev mode extended along the wire, and (c) a scenario of two pairs of quasi-Majorana modes~\cite{Prada12, Kells12, Liu17, Vuik19}. Note that scenario (b) is continuously connected to (a) as the non-local tunnel couplings are progressively reduced. 

To gain intuition, it's instructive to first consider these single-wire scenarios embedded in a $Z$-loop interferometer. Using the convention depicted in \Cref{fig:alt_scen} and performing a calculation analogous to \Cref{app:model} yields an off-diagonal component of the Hamiltonian (describing the coupling of the readout QD to the wire) of the form 
\begin{align}
t_C(\varphi,Z)=t_{A1}+t_{B1}e^{i\varphi}+i(t_{A2}+t_{B2}e^{i\varphi})Z,
\end{align}
where $\varphi$ is determined by the flux through the interferometer loop and $Z=i\gamma_1\gamma_2$ is the fermion parity. The corresponding QD-wire energy spectrum reads
\begin{align}
E(\varphi,Z)=\pm \frac{1}{2}\sqrt{(E_D+2 Z E_{12})^2+|t_C(\varphi, Z)|^2}
\end{align}
with $E_D$ and $E_{12}$ being QD detuning and MZM splitting energy, respectively. One can observe that when $E_{12} = 0$, the topological scenario
($t_{B1} = t_{A2} = 0$)
is symmetric under the transformation $\varphi \mapsto \varphi+\pi, Z\mapsto -Z$, which ensures $h/2e$ flux periodicity of $|\dCq|$. By contrast, finite non-local couplings $t_{B1},t_{A2}$ break this symmetry and generally lead to an $h/e$-periodic response; this, in turn, allows one to rule out the Andreev‑state scenario (b) on general grounds. Importantly, the $h/e$ periodicity arising from non‑local couplings at $E_{12}=0$ is distinct from the behavior at finite $E_{12}$, which yields a characteristic zig‑zag structure in flux–QD‑detuning maps~\cite{Aghaee25}. In the former case, the $\dCq$ response remains centered around $E_D = 0$ and exhibits even/odd oscillations in the amplitude of $|\dCq|$, leading overall to $h/e$ periodicity. Finally, scenario (c) was considered in Ref.~\onlinecite{Aghaee25} which concluded that it's generally incompatible with the long parity-switching time scales observed in the $Z$ measurement, see also below. 

Let us now consider these scenarios starting with the bottom wire, which is part of the $Z$ measurement loop and is tuned by WP2.
First, we note that
the absence of a zig-zag pattern as a function of $\Delta V_\mathrm{QDL}$ in \Cref{fig:zmpr} requires $E_{12}$ to be smaller than the resolution of the measurement $\lesssim \SI{1}{\micro\electronvolt}$.
Next, we observe that
the data presented
in \Cref{fig:zmpr} strongly constrains
scenarios (b) and (c).
The clear $h/2e$ periodicity (\Cref{it:Zflux}) indicates that the
non-local couplings are small: $t_{A2},t_{B1}\ll t_{A1},t_{B2}$ .
This is, in fact, expected from
the TGP data (\Cref{it:tgp}) since
the observed ZBPs at the left end
require large local
couplings $t_{A1},t_{B2}$,
which, in turn, implies
the non-local couplings
$t_{A2},t_{B1}$ are small,
given the length
of the wire. Thus scenario (b) is
unlikely.
Observing ZBPs on both sides of the wire is more natural in scenario (c). However, following the discussion in \cite{Aghaee25}, in order to reproduce the observed $\tau_Z$ time scales, the energies of the local modes would need to be fine tuned to the $\SI{}{\nano\electronvolt}$ scale.  In the absence of that, extra Majorana modes must be gapped
out via $\tilde{E}_{12}$ in order for bi-modality to be resolvable. The size of $\dCq$ requires this gap to be of the order of temperature, which essentially recovers the topological scenario (a).

We now consider the top
wire segment, which is part of the $X$ measurement loop
and is tuned by WP1. Starting with scenario (c), the faster $\tau_X$ time scales put a weaker bound (compared to the lower wire) on the level of fine-tuning $< \SI{0.3}{\micro\electronvolt}$ (same as our estimate for residual $E_M$) of the local state in the upper wire segment that is involved in the $X$ measurement loop. This small energy splitting could change sign and thus lead to a similar second time scale as observed in \Cref{it:2times} due to two processes: (i) QP poisoning events, (ii) parity exchange with the other pair of quasi-Majorana modes (i.e. the local mode at the other end of the top wire). Given the long time scales, this again puts a $\SI{}{\nano\electronvolt}$ scale bound on the coupling between the local states, $\tilde{E}_{12}$ in \Cref{fig:alt_scen}(c). This quasi-Majorna scenario is therefore in principle possible but requires extremely well-separated local modes that are tuned to low (sub-$\SI{}{\micro\electronvolt}$) energies.

Finally, we analyze a configuration that combines scenario (a) in the lower wire with scenario (b) in the upper wire; see \Cref{fig:alt_scen}(d). The total Hamiltonian is
\begin{align}
H = H_M + H_1,
\label{eq:H1}
\end{align}
where $H_M$ is defined in \eqref{eq:HM} and $H_1$ is
\begin{widetext}
\begin{align}
H_1 =
\left(
\begin{array}{cccc}
    0 & 0 & 0 & t_\X^{(2)} - i M t_\X^{(4)} e^{-i\varphi} \\
    0 & 0 & -t_X^{(2)*} + i M t_X^{(4)*} e^{i\varphi} & 0 \\
    0 & -t_\X^{(2)} - i M t_\X^{(4)} e^{-i\varphi} & 0 & 0 \\
    t_X^{(2)*} + i M t_X^{(4)*} e^{i\varphi} & 0 & 0 & 0 \\
\end{array}
\right)\!.
\end{align}
\end{widetext}
Here $t_\X^{(2)}$ and $t_\X^{(4)}$ denote the couplings between the QD and the outer MZMs. The spectrum of $H$ is invariant under $\varphi \mapsto \varphi + \pi$, implying that $|\dCq|$ is $h/2e$-periodic. If the backbone connecting the two wires were not fully depleted, additional tunnel couplings connecting the MZMs and QD across the backbone would arise; these generally break the $h/2e$ periodicity, in a manner analogous to a $Z$ measurement of scenario (b). Thus, the experimentally observed $h/2e$ periodicity provides further evidence that the trivial backbone is insulating.

As discussed above, the $Z$-measurement imposes a stringent constraint on the Hamiltonian $H$ \eqref{eq:H1}, specifically requiring $t_\X^{(4)}\ll t_\X^{(3)}$. This condition leads to scenario (d) illustrated in \Cref{fig:alt_scen}. However, we cannot similarly constrain $t_\X^{(2)}$ because this device design does not enable a $Z$ measurement of the top wire. Moreover, the X measurement described by \Cref{eq:H1} generally yields an $h/2e$-periodic response so it doesn't allow us to constrain $t_\X^{(2)}$ based on the observed flux periodicity. Bounds on $t_\X^{(2)}$ thus depend on the details of the system including the noise channels coupling to the system.
Nevertheless, the presence of ZBPs at opposite ends of the top wire, combined with its length of $\SI{3.5}{\micro\meter}$, supports the assumption of a weak non-local coupling $t_\X^{(2)}\ll t_\X^{(1)}$, which is consistent with the observed data. 

In conclusion, we interpret our measurements in terms of $X$ and $Z$ loop measurements of a tetron (with small but finite intra-wire coupling between the Majorana modes). As described above, alternative scenarios may reproduce some of the experimental observations. The entirety of the data, however, strongly constrains alternative scenarios and is naturally explained by the topological interpretation.

\section{Error model} \label{app:error_model}

Similar to the operational assignment error for a series of qubit measurements \cite{Aasen25}, we can define an equivalent metric for the continuous $X$ and $Z$ measurements that we perform here via
\begin{equation}
    \erra^\mathcal{O}=\text{max}_{r,s}\left| P(\mathcal{M}^\mathcal{O}_r|\mathcal{M}^\mathcal{O}_s)-\delta_{r,s}\right|,
\end{equation}
where $r,s =\pm 1$ are the two observed measurement outcomes for each of the measurements $\mathcal{M^O}$ with $\mathcal{O}=X,Z$. 
Here, following the notation of Ref.~\onlinecite{Aasen25}, we use $\mathcal{M^O}$ to denote the noisy, imperfect implementation of the projective measurement $\mathcal{O}$. 
Note that here we are continuously measuring so the distinction between different measurements is done at the level of the analysis by interpreting the collected data points as separate measurements.

A simple qubit error model can be obtained by taking into account the lifetime $\tau$ and assignment errors due to finite SNR independently. 
This yields
\begin{equation} \label{eqn:erra-expression}
    \erra = \frac{1}{2}\left[ 1-e^{-2t/\tau}\erf(\SNR/\sqrt{2})^2\right],
\end{equation}
where the squares arise from the possibility of errors appearing in either one of the measurements entering $\erra$. 
The parameters of this model can be fixed by corresponding lifetimes extracted from the autocorrelation function ($\tau = \tauX$) or dwell time histograms ($\tau = \tauZ$) together with the extracted SNR from fitting a Gaussian Mixture Model (GMM) of the time records show in \Cref{fig:xmpr} and \Cref{fig:zmpr}. 
Using the expected scaling of $\text{SNR}\propto \sqrt{\tau_m}$, where $\tau_m$ is the integration or measurement time, we then compare the theoretical model to the experimentally extracted $\erra$ when applying different coarsening of the underlying data points. 
The good agreement found in \Cref{fig:error_metrics} indicates that the error model captures the main effects leading to the observed assignment errors.  

\section{Device design} \label{app:design}

\begin{figure}
\centering
\includegraphics[width=\columnwidth]{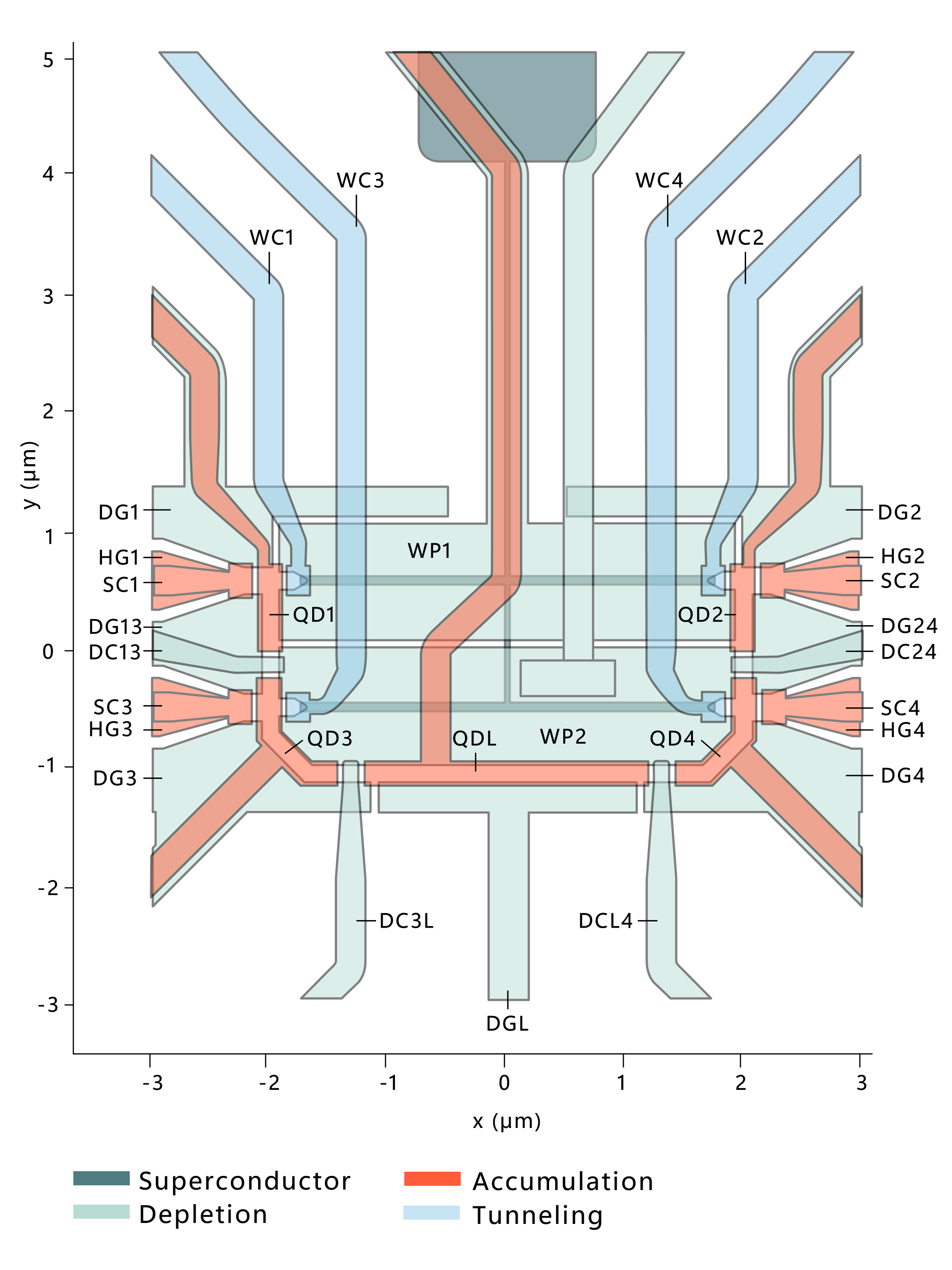}
\caption{
Active area of a tetron qubit device with the gate layout colored according the tuning of the underlying 2DEG.  
Light green indicates full depletion of the underlying 2DEG, red indicates accumulation of the underlying 2DEG, and blue indicates tuning the underlying junction region to the tunneling regime. 
The device consists of a 2DEG stack, superconductor (dark green), and two layers of gates separated by dielectric. The area surrounding the device is depleted by the depletion gates ``DGi.'' 
Plunger gates WP1, WP2 simultaneously deplete the exposed 2DEG, tune the horizontal nanowire segments into their lowest electric subband, and fully deplete the narrower vertical superconducting backbone. 
Helper gates (red, HGi) are set to accumulation and are connected to Ohmics outside of the active area to provide a 2DEG reservoir. 
Quantum dot plungers are set to accumulate electrons in the respective quantum dots. 
Cutter gates are set to form the necessary tunnel junctions between dots, wires and transport leads; the configuration illustrated here is used for transport-based tune-up steps, see \Cref{app:device_tuning}.
}
\label{fig:tetron_gates}
\end{figure}

The device shown in \Cref{fig:tetron_gates} represents a specific implementation of the tetron qubit~\cite{Karzig17, Knapp18b}. 
The material stack is similar to that described in Ref.~\onlinecite{Aghaee25}, with two key differences the Al nanowire is now $\SI{2.5}{\nano\meter}$ thick and the intermediate AlO${}_x$ layer between the semiconductor and HfO${}_x$ gate dielectric has been removed. 
The InAs quantum well, wire width, two layers of dielectrics, and gates remain unchanged. 
There are two parallel $\SI{3.5}{\micro\meter}$ long, $\SI{60}{\nano\meter}$ wide Al strips, epitaxially grown on the semiconductor, and $\SI{1}{\micro\meter}$ apart.
These wires are connected to each other and the $3.0 \times \SI{1.5}{\micro\meter}^2$ Al pad with a $\SI{40}{\nano\meter}$ wide Al ``backbone.'' 
The Al pad is $\SI{3.5}{\micro\meter}$ away from the top wire and is connected to ground by a low-resistance Ohmic contact~\cite{Aghaee25}. 
The electron density in the quantum well in the area surrounding the superconducting wires is controlled by the first layer plunger gates WP1, WP2: conducting channels are formed under two parallel superconducting wires and the quantum well is depleted elsewhere. 
The 2DEG surrounding the device is depleted by 7 depletion gates ``DGi'' in \Cref{fig:tetron_gates}. 
Helper gates ``HGi'' run from the qubit region all the way to metallic Ohmic source contacts outside the field of view of \Cref{fig:tetron_gates} but near the edge of the 2DEG mesa and provide reservoirs for the 2DEG density next to the device for DC transport experiments.
There are five quantum dots in the device (QD1, QD2, QD3, QD4, QDL) which are laterally confined by the plunger and depletion gates, which are separated by $\SI{150}{\nano\meter}$. 
Dots 1 and 2 are $\SI{0.745}{\micro\meter}$ long
and are perpendicular to the wires. 
Dots $3$ and $4$ are curved, with total length of $\SI{1.2}{\micro\meter}$, and the ``long'' dot is parallel to the SC wires and is $\SI{2.4}{\micro\meter}$ long. 

The quantum dots are covered with corresponding dot ``plunger'' QDi gates in the second layer.
They set the electrical potential and 2DEG density in each dot. 
The junctions between adjacent dots, nanowires and ohmics are controlled by the cutter gates. 
Quantum dot cutter ``DCij'' gates control inter-dot tunnel couplings; wire cutters ``WCi'' control the coupling between the dots and the nanowires; and source cutter ``SCi'' gates control the coupling between the small dots and the two-dimensional electron gas (2DEG) reservoirs. 
In the tuning stage, the source cutters are opened in order to enable transport measurements, after which the source cutters are closed to decouple the system from the leads. 

\begin{figure}
\includegraphics[width=\columnwidth]{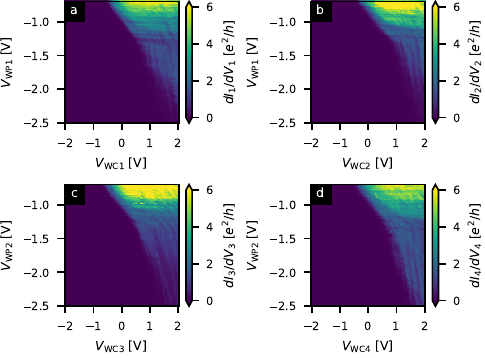}
\caption{The differential local conductances, $dI_j/dV_j$, at the four ends of the two wires in the tetron device discussed in the main text, measured at $\SI{0.5}{\milli\volt}$ bias as a function of the junction cutter and wire plunger gates. Discrete conductance steps are visible, demonstrating the quality of the junctions.} 
\label{fig:junctions}
\end{figure}

The tetron design incorporating a backbone introduces several key differences compared to the linear design used in Ref.~\onlinecite{Aghaee25}. 
First, the nanowires are terminated at junctions, enabling a simplified and more robust junction tuning than the previously employed side junctions. 
These junctions have been optimized for stronger coupling to wire states and more robustness with respect to disorder, as evidenced by the conductance plateaus observed as a function of the wire plunger gate, shown in \Cref{fig:junctions}. 
Second, the device design is scalable to arrays of topological qubits~\cite{Karzig17, Aasen25}. 
Third, the two-dimensional geometry of the device leads to a more compact footprint compared to the linear device. 
This results in smaller capacitive losses in readout lines and more optimal RF performance. 

\begin{figure}
\centering
\includegraphics[width=\columnwidth]{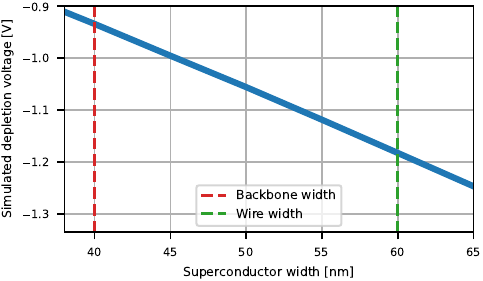}
\caption{
Simulation of the depletion voltage of a wire for varying superconductor width. 
The depletion voltage is calculated as the onset in gate voltage of the lowest subband from a cross-section of the device as described in Refs.~\onlinecite{Antipov18, Winkler19}. 
In a large voltage range above the depletion voltage of the wires, comprising the lowest subband, the backbone is depleted.
}
\label{fig:depletion_voltage_width}
\end{figure}

The improvements in the tetron design come at the cost of adding the backbone, which introduces several considerations to the design.
The 2DEG under the backbone needs to be fully depleted to avoid low energy excitations under the backbone that couple the two nanowires. 
The superconductor screens the electric field from the plunger gates, so the density of electrons under the superconductor is controlled by the lateral confinement from the electric field coming from the sides. The confinement depends on the width: the wider the wire, the more negative its depletion voltage. 
By making the backbone $\SI{20}{\nano\meter}$ narrower than the wires, the semiconductor under the backbone will be fully depleted when the wires themselves are tuned into the lowest subband.
We have verified this in simulations, see \Cref{fig:depletion_voltage_width}.

\section{Readout system} \label{app:readout_system}

\paragraph{Dispersive gate sensing.}

Fermion parity is measured via dispersive gate sensing~\cite{Colless13}.  
The method is similar to that of Ref.~\onlinecite{Aghaee25}: an off-chip spiral inductor is connected to the plunger gate of a quantum dot to create a high-impedance resonant mode that can be used to sense the quantum capacitance of the dot.  
When the quantum dot is incorporated into an interference loop, the quantum capacitance of the dot becomes parity dependent.  
Microwave reflectometry then permits the detection of small parity-dependent changes in the resonant frequency. 

\begin{figure}
\includegraphics[width=0.95\columnwidth]{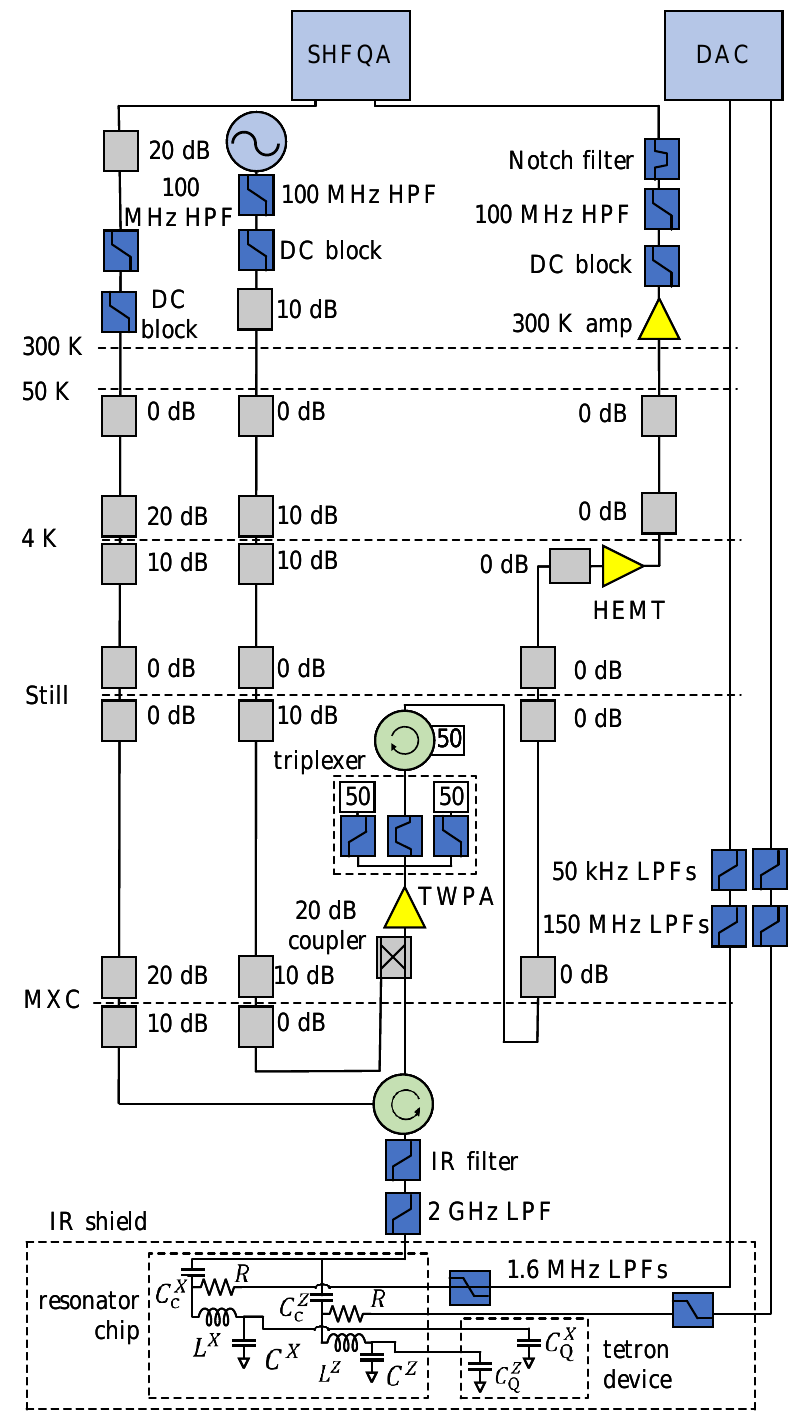}
\caption{
Wiring diagram for the readout system. 
The quantum dots used for dispersive gate sensing  appear in the circuit schematic as parity-dependent quantum capacitances $\Cq$ (bottom right, in the box marked ``tetron device'').  
Parameters for the circuit components are given in \Cref{table:readout_circuit_params}.
}
\label{fig:wiring_diagram}
\end{figure}

The readout system used for these measurements is depicted schematically in~\Cref{fig:wiring_diagram}. 
The circuit diagram at the bottom of the schematic illustrates how the readout circuitry for both the $X$ and $Z$ loops are frequency-domain-multiplexed on a single readout lane.  
Readout circuitry for the other $X$ loop (connected to quantum dot 2) is also present on the device, but is omitted from the schematic for clarity.  
Parameters for the circuit components and the readout modes are given in~\Cref{table:readout_circuit_params}.

\begin{table}[ht!]
\centering
\begin{tabular}{|c|c|c|c|c|c|c|c|} 
 \hline
 Loop & Quantum dot & $L$ [nH]& $C$ [fF]& $C_\mathrm{c}$ [fF]& $f_0$ [MHz]& $Q_\mathrm{c}$ & $Q_\mathrm{i}$ \\ [0.5ex] 
 \hline\hline
 $Z$ & QDL
 & 247  
 & 197 & 910 & 722 & 36 & 160 \\ \hline 
 $X$ & QD1 
 & 173 
 & 224 & 840 & 808 & 25 & 145\\ \hline 
\end{tabular}
\caption{Circuit parameters for the readout circuit depicted in~\Cref{fig:wiring_diagram} at \SI{2.3}{\tesla}.  The $R = 20$~k$\Omega$ bias tee resistors are the same for both loops. The ``Quantum dot'' column indicates which dot is utilized for dispersive gate sensing. The value of $\Cq$ depends on the device tuning and readout drive strength but reaches up to $\dCq^\Z \approx \SI{600}{\atto\farad}$ and $\dCq^\X \approx \SI{300}{\atto\farad}$. 
} 
\label{table:readout_circuit_params}
\end{table}

A second readout lane, identical to the one illustrated in~\Cref{fig:wiring_diagram} but without the JTWPA, triplexer, and directional coupler, is used to multiplex the dispersive gate sensing of quantum dots 3 and 4. 
Measurements on these dots are used for tuning of the device. 
\Cref{table:unused_readout_circuit_params} shows the values of the inductors and the resonance frequencies for the readout resonators used for dispersive gate sensing of quantum dots 2, 3, and 4.

\begin{table}[ht!]
\centering
\begin{tabular}{|c|c|c|} 
 \hline
Quantum dot & $L$ [nH] & $f_0$ [MHz]\\ [0.5ex] 
 \hline\hline
 2 & 278 & 656 \\ \hline 
 3 & 235  & 692 \\ \hline  
 4 & 185  & 755   \\ \hline 
\end{tabular}
\caption{Resonant frequencies and inductances for the backup $X$ loop (quantum dot 2) and the readout circuits used for bringup (quantum dots 3 and 4).}
\label{table:unused_readout_circuit_params}
\end{table}

To increase SNR, a JTWPA~\cite{Macklin2015} is used as the first amplifier in the readout chain.  
A triplexer is inserted at the output of the JTWPA to improve the impedance matching outside of the operating bandwidth of the subsequent isolator. 

\paragraph{Diversity combining.}

The finite gain of the JTWPA in our setup does not entirely overwhelm the added noise of downstream components in the readout chain. 
To recover some part of this lost SNR, we measure both the signal and the idler tones and diversity combine~\cite{Brennan03} the results.

Diversity combining is possible because, as in any parametric amplifier, the mixing process that amplifies the signal also creates an entangled photon at the idler frequency. 
The JTWPA used in our setup utilizes the nonlinear inductance of the Josephson junctions to amplify the readout signal using a degenerate four-wave mixing process. This coupling of a strong pump (at $\omega_p$) and a signal tone ($\omega_s$) gives rise to an intermodulation product, the idler tone ($\omega_i$), obeying $\omega_s + \omega_i = 2 \omega_p$.  This mixing process converts two frequency-degenerate pump photons into a pair of entangled signal and idler photons~\cite{Qiu23}. 
A solution to the coupled wave equations, in the limit of small signal and an undepleted pump, relates the amplitude of the coupled signal ($a_s$) and idler ($a_i$) waves~\cite{OBrien14}:
\begin{equation}
    \dfrac{\partial}{\partial x} \begin{pmatrix} a_s\\a_i \end{pmatrix}
    \propto \begin{pmatrix} a_i^\dagger\\a_s^\dagger \end{pmatrix} 
    e^{i\Delta k x}.
    \label{eq:parametric_amp}
\end{equation}
Here $x$ is the spatial coordinate along the transmission line formed by the JTWPA and $\Delta k$ is a constant related to the signal, pump, and idler wavevectors, and other constants. 
\Cref{eq:parametric_amp} is symmetric with respect to exchange of the signal and idler amplitudes up to prefactors. 
This symmetry allows the idler to carry the same information as the signal when the nonlinear transmission line is long.

Diversity combining is beneficial because the noise added by components downstream of the JTWPA is uncorrelated at the signal and idler frequencies. (And because the JTWPA gain is finite --- in the limit of infinite gain there would be no benefit from diversity combining). 
In the opposite limit where the system noise at signal and idler frequencies is uncorrelated and the noise at the signal and idler frequencies is equal, diversity combining improves SNR by a factor of $\sqrt{2}$.

Measuring the idler enables two simultaneous measurements of the device response with partially different noise realizations. 
The idler response is then mapped back onto the signal by conjugating and rescaling it to account for the background experienced at different readout frequencies. 
Finally, in post-processing we diversity combine the two time-resolved measurements at the signal and idler frequencies by summing the two tones weighted by their respective SNR. 
This method, known as maximal-ratio diversity combining, is optimal under reasonable assumptions~\cite{Brennan03}.

\paragraph{Digital filtering.} \label{app:digital_filtering}

The demodulated readout waveform typically contains power at microwave frequencies other than the tone employed for reflectometry.  Such spurious peaks arise from two sources: i. unwanted mixing products in the JTWPA and ii. from the pickup of low-frequency coherent signals on the plunger gates which may then be transduced into sidebands on the readout tone. 

To suppress the effect of these spurious peaks we utilize two mitigations.  First, we employ a window function with $\SI{90}{\nano\second}$ cosine edges. Second, we digitally filter the readout signal with 1st-order Bessel infinite impulse response (IIR) notch filters centered at the spurious peaks in the readout spectrum.

\section{Device tuning} \label{app:device_tuning}

\paragraph{Transport measurements}
First, with all gates set to allow for transport measurements as shown in \Cref{fig:tetron_gates} (DCij, DGi in depletion, QDi, SCi, Hi in accumulation and WCi in the tunneling regime) we characterize the depletion voltages of NWs and backbone, as well as induced gap, and parent gap of both wire segments using local and non-local conductance spectroscopy performed on all four terminals of the device, using methods described in Refs.~\onlinecite{Aghaee23, Aghaee25}. 

\begin{figure*}
\centering
\includegraphics[width=\textwidth]{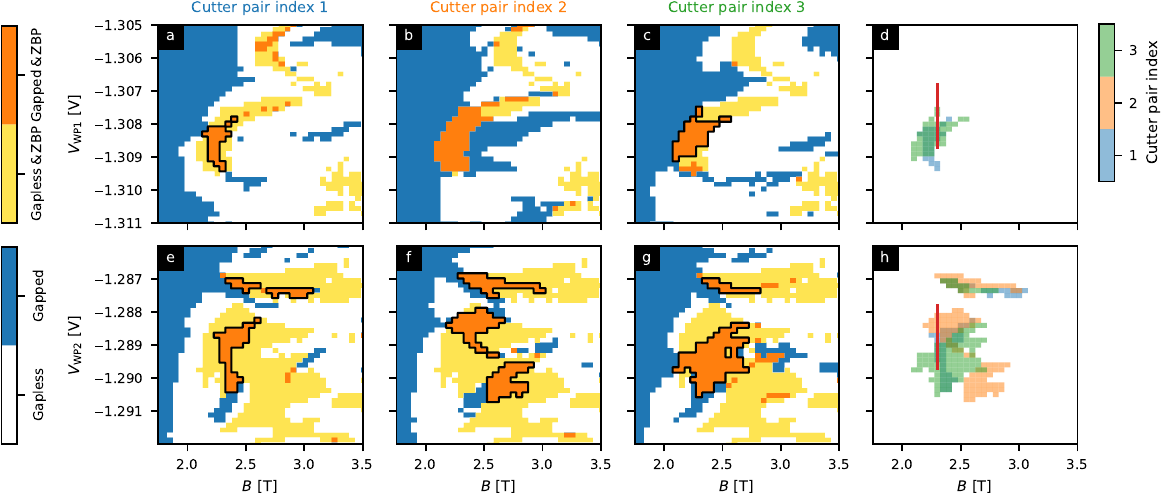}
\caption{
Experimental TGP phase diagram of the device measured prior to the $X$ loop configuration. 
Panels (a-c) show phase diagrams of the top wire for the different cutter settings. Solid lines show the subregions of interest for the cutter settings. 
Panel~(d) shows the TGP region of interest as deduced by checking the overlap of subregions of interest from different cutters. 
Panels~(e-h) show analogous phase diagrams for the bottom wire. 
For tuning the $X$ loop additional transport scans were performed, shown in \Cref{fig:transport_x_loop}. 
The range of the smaller scan in \Cref{fig:transport_x_loop} is shown in (d) and (h) as red lines.
}
\label{fig:tgp_phase_diagram}
\end{figure*}

In a $\approx \SI{100}{\milli\volt}$ wire plunger voltage range above wire depletion we run stage 1 and stage 2 of the TGP ~\cite{Aghaee23, Aghaee25} to identify suitable clusters for the parity measurements described in the main text. The results of stage 2 of TGP for the clusters used in the experiments are shown in \Cref{fig:tgp_phase_diagram}. 
For all subsequent measurements the in-plane magnetic field is set to $\SI{2.3}{\tesla}$, where the clusters overlap. Having an in-plane magnetic field selected, and suitable $V_\mathrm{WP1}$ and $V_\mathrm{WP2}$ ranges identified, the gates SCi are set to depletion ($\mathrm{SCi} = \SI{-1.25}{\volt}$) and the nanowires are disconnected from the leads. 
Subsequently we shift to RF measurements and calibrate the resonance frequency of the resonators attached to QD1 and QD3 in this tuning configuration. 
We find resonance frequencies of $f_\mathrm{QD_1} = \SI{808}{\mega\hertz}$ and $f_\mathrm{QD_3} = \SI{692}{\mega\hertz}$. 
We optimize  the pump power and frequency of the Josephson Travelling Wave Parametric Amplifier (JTWPA) at the high-SNR readout frequency $f_{QD_1}$. 
 
\begin{figure}
    \centering
    \includegraphics[width=\columnwidth]{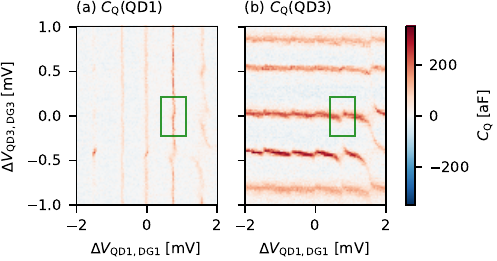}
    \includegraphics[width=\columnwidth]{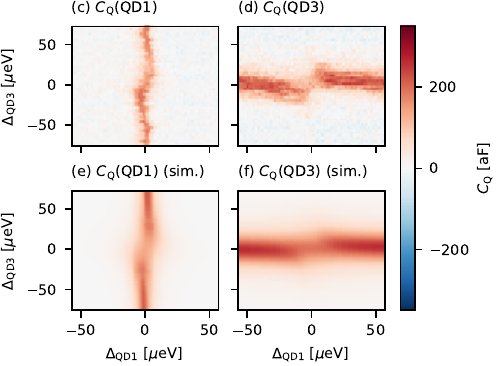}
    \caption{
    (a,b)~Measured quantum capacitance response of dot 1 and dot 3 used for $X$ loop interferometer balancing. Gates $\mathrm{QD}j$ and $\mathrm{DG}j$ are swept together ($j=1,3$). 
    (c,d)~Zoom-in on the charge transitions used to obtain the results of \Cref{fig:xmpr}. The zoom-in area is indicated by green boxes in panel (a,b). (e,f)~Simulation results for the best fit parameters (see main text).
}
    \label{fig:qdmzm}
\end{figure}

\paragraph{$X$ loop interferometer balancing.}

Similarly to Ref.~\onlinecite{Aghaee25}, the arms of the interferometer must be balanced in order to maximize the measured signal. The $X$ loop of the tetron device is formed using two quantum dots, where dot 1 is used for readout while dot 3 mediates an effective coupling between dot 1 and the bottom wire.
The gate DC13 is used to control the interdot coupling $t_{13}$ while gate WC1 (WC3) controls the coupling $t_{m1}$ ($t_{m3}$) between dot 1(3) and the neighboring wire. These gate voltages must be selected such that the direct coupling $t_{m1} \equiv t_\X^{(1)}$ is balanced with the effective coupling $t_\X^{(3)}$, where $t_\X^{(3)} \sim t_{m3} t_{13}/\Delta_3$ is the effective coupling mediated through dot 3 and $\Delta_3 = E_{C3} (1-2n_{g3})$ is the detuning of dot 3 controlled using QD3 with $n_{g3}$ the dot gate charge. 

To achieve this, we measure the response of resonators coupled to QD1 and QD3 as their plunger gates are swept across roughly 4-5 charge transitions. The resonator responses are then converted into $\Cq$ signals~\cite{Aghaee25} which are analyzed as described below to estimate the couplings.
The tunnel couplings can change from one transition to another, offering an extra tuning knob for achieving the correct tunnel couplings in addition to the gates WC1, WC3, and DC13.
In addition to the desired $\Cq$ signals, varying $V_\mathrm{QD1(3)}$ can change parasitic couplings along the gate routes leading to a voltage-dependent background in the measurement. To mitigate this effect, we sweep QD1(3) together with the first layer gate routed under it: DG1(3).

\Cref{fig:qdmzm}(a,b) shows the measurement used to characterize couplings in the same gate configuration as \Cref{fig:xmpr}, with panels (c,d) a zoom-in on the charge transitions selected for the following $X$ loop parity readout measurements.
To estimate the couplings, we fit the measured data to reference simulations of the dynamical quantum capacitance. The reference simulation data is generated using the open systems dynamics methods described in Refs.~\cite{Boutin25, Aghaee25}. From this fit procedure, we estimate $t_{m1}\simeq\SI{1.6}{\ueV}$, $t_{m3}\simeq\SI{9.2}{\ueV}$ and $t_{13}\simeq\SI{10}{\ueV}$. 
\Cref{fig:qdmzm}(e,f) show simulation results based on the above parameters on the same scale as the above panels (c,d). Lever arms $\alpha_\mathrm{QD1}=0.15$ and $\alpha_\mathrm{QD3}=0.25$ both convert the measured voltages to detunings $\Delta_\mathrm{QD1}$ and $\Delta_\mathrm{QD3}$. The gate DG1(3) also has a lever arm to dot 1(3) that must be taken into account when converting from voltage to detuning energy. The ratio of their lever arms $\alpha_\mathrm{DG1}/\alpha_\mathrm{QD1}=0.1$, $\alpha_\mathrm{DG3}/\alpha_\mathrm{QD3}=0.35$ are extracted from the slope of charge transitions as QD1(3) and DG1(3) are swept separately.

We measure Coulomb diamonds in transport in a similar device and extract a mean addition energy for dot~3 $E_{C3}=\SI{120}{\ueV}$. In the loop configuration, the addition energy will be reduced compared to the transport configuration by quantum charge fluctuations~\cite{Flensberg1993}. The renormalized addition energy $E_{C3}'$ can be estimated either numerically~\cite{Boutin25} or in perturbation theory by considering the coupling of the dot 3 level nearest charge degeneracy to additional levels in dot 1 and to states in the neighboring wire at the topological gap edge. Using the latter technique, we estimate $E_{C3}'\sim 0.65 E_{C3}$. From these parameters, we evaluate the effective coupling $t_\X^{(3)}$ numerically as a function of $n_{g3}$. Since $t_{m3},t_{13}\gg t_{m1}$, a balanced interferometer is achieved in a regime where $n_{g3}\sim 0.15$, in good agreement with the regime of maximal signal observed in \Cref{fig:xmpr}(b). 

\begin{figure}
\includegraphics[width=\columnwidth]{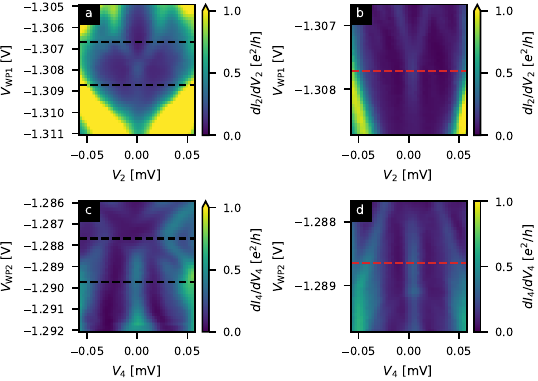}
\caption{
Comparison of transport data at junctions 2 and 4, measured during TGP (a,c) and in the $X$ loop configuration (b,d) before parity measurements. The black dashed lines in (a,c) indicate the range scanned in (b,d) and the red dashed lines indicate the $V_\mathrm{WP1}$ and $V_\mathrm{WP2}$ values used for the initial interferometric measurements. These values were selected approximately at the center of the zero bias peaks, as visible also from the intensity of those peaks. We also note a small $V_\mathrm{WP}$ shift in zero bias peak location between TGP and $X$ loop configurations.}
\label{fig:transport_x_loop}
\end{figure}

With the interferometer properly tuned, we can monitor the regions identified by the TGP through transport measurements using junctions 2 and 4, temporarily setting gates SC2 and SC4 to accumulation. This enables precise adjustment of  $V_\mathrm{WP1}$ and $V_\mathrm{WP2}$ without disturbing the interferometer. 
\Cref{fig:transport_x_loop} presents a comparison between the transport data collected during TGP and the data from the opposite junctions of the interferometer prior to the parity readout measurements.

\begin{figure}
    \centering
    \includegraphics[width=0.7\linewidth]{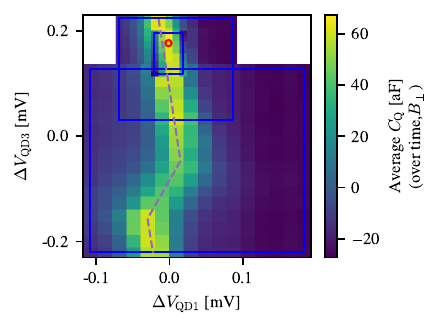}
    \caption{Average of the $\Cq$ response over time and $B_\perp$ using the same underlying data as \Cref{fig:xmpr}(b-h) but including regions outside the plot range of \Cref{fig:xmpr}(b). The purple dashed line is the same guide to the eye of \Cref{fig:xmpr}(b) indicating the QD1 charge resonance with QD3 (positive slope) and the nanowire (negative slope), while the red circle marker denotes the gate voltages used in panels \Cref{fig:xmpr}(c-h). The small jumps in the data are due to the plot showing 3 different measurement datasets (data ranges marked by blue boxes) consecutively zooming in to the regime of maximal visibility for the $X$ measurement.} 
    \label{fig:average_qd_qd}
\end{figure}

\paragraph{$X$ loop parity readout.}

After identifying the dot 1 and dot 3 transitions with the desired tunnel couplings, and optimizing the rf drive power, we collect time records of $\Cq$. The core measurement involves a multi-dimensional parameter sweep, where $\Cq$ time records are taken as functions of out-of-plane magnetic field, QD3 plunger voltage and QD1 plunger voltage --- in order of the outer-most to inner-most parameter of the measurement loop. 
For all time records, we discard the initial \SI{200}{\micro\second} of data to account for settling time of stepped DC voltages through low-pass filters internal to the DAC.
The magnetic field axis is used to analyze the flux dependence of the signal, while adjusting QD1 and QD3 plunger voltages allows precise tuning of the interferometer balancing.

\Cref{fig:xmpr}(b) shows the excess kurtosis of measured $\Cq$ as a function of QD1 and QD3. In \Cref{fig:average_qd_qd} we show the average of $\Cq$ over time and $B_\perp$ for the same dataset to show the lines of charge resonance of QD1 with QD3 and the nanowire. \Cref{fig:average_qd_qd} is the aggregation of three different measured datasets following the parameter sweep described above, with each consecutive dataset increasing the $V_{\rm QD1}$ and $V_{\rm QD3}$ resolution in the region of maximal signal. The finer resolution dataset was obtained following the drive amplitude optimization (see next section) and uses a larger drive amplitude compared to the coarser scans. As shown by the red circle in \Cref{fig:xmpr}(b), the flux-periodic bimodal response is limited to the regime close to QD1-nanowire resonance with appropriate QD3 detuning to balance out the arms of the interferometer.

To perform interferometric parity measurements as functions of $V_\mathrm{WP1}$ and $V_\mathrm{WP2}$, we compensate for the finite cross capacitance between these gates and dot 1 and dot 3. At each point in $V_\mathrm{WP1}$ and $V_\mathrm{WP2}$, image convolution analysis is used to  properly calibrate the dot detunings, enabling automated exploration of the parameter space.   

\begin{figure}
\includegraphics[width=\columnwidth]{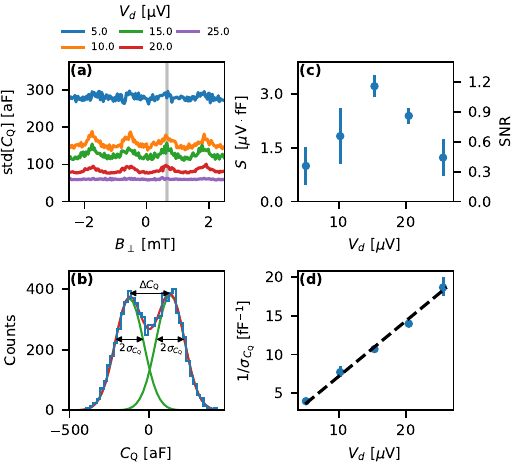}
\caption{Optimization of readout tone amplitude $V_d$.
(a)~Standard deviation of $\Cq$ time records 
for fixed $V_{\rm QD1}$ and $V_{\rm QD3}$
corresponding to the red circle in \Cref{fig:average_qd_qd} and \Cref{fig:xmpr}(b).
(b)~Example of the SNR analysis based on a bimodal Gaussian mixture fit to the histogram of time records.  
(c)~Signal $S = V_d \dCq$ extracted by repeating the analysis in panel (b) over time records collected in the shaded area indicated in panel (a) and over 8 similar datasets where QD3 was scanned over a range $\Delta V_{\rm QD3} = \SI{80}{\uV}$. The blue disks indicate the average of the fit parameters and the error bar the standard deviation.
(d)~Inverse of the Gaussian width $\sigma_{\Cq}$ as a function of $V_d$. The black dashed line indicate a linear fit extracting $\sigma_R=2.8 \pm 0.2$.
}
\label{fig:tuning_drive_sweep}
\end{figure}

\paragraph{RF drive amplitude.} \label{app:drive_sweep}

Once a signal is identified, we perform the $X$ loop parity readout for several readout tone amplitude $V_d$ values with the goal of maximizing SNR. For small $V_d$, we expect SNR to improve linearly with $V_d$. As $V_d$ increases backaction effects are expected to reduce the measured quantum capacitance $\dCq$ leading to an optimal amplitude maximizing SNR~\cite{Boutin25}.
As described in \Cref{app:readout_system}, there is significant attenuation between the room temperature generator and the device at cryogenic temperatures. We calibrate the 
drive amplitude on QD1 such that $V_d = \lambda_{\rm att} V_d^{\rm RT}$, where $V_d^{\rm RT}$ is the drive amplitude at the generator output and $\lambda_{\rm att}$ is an attenuation coefficient. The attenuation is characterized by an RF measurement of dot 1 as a function of QD1 and $V_d^{\rm RT}$ in a regime where dot 1 is weakly connected to the transport lead~\cite{Koski18}. Using a linear fit of the broadening of the RF response with $V_d^{\rm RT}$ allows us to calibrate $\lambda_{\rm att}=(2.55\pm 0.2) \times 10^{-3}$.

\Cref{fig:tuning_drive_sweep} presents the data of \Cref{fig:xmpr} of the main text together with additional data for varying $V_d$. Panel (a) shows the standard deviation of time records as a function $\Bperp$ and $V_d$. Three $\Bperp$ values  (indicated by a shaded gray area) where the standard deviation reaches a local maxima are selected to study and maximize SNR. For each time records, a fit of a bimodal Gaussian mixture is performed on the $\Cq$ histogram to extract the distance between the modes $\dCq$ and their standard deviations $\sigma_{\Cq}$. See panel (b) for an example fit. To gather statistics and find an optimal drive amplitude valid over a range of tuning parameters, the fit is also repeated for 8 different $V_{\mathrm{QD}3}$ spanning a range of $\SI{80}{\uV}$.
As the measured quantum capacitance $\dCq$ can decrease with $V_d$ due to backaction effects, this is not directly the quantity to optimize in order to maximize SNR. Instead, we parametrize the measurement SNR as ${\rm SNR} = S\sqrt{\tau_m}/\sigma_R$ where we define the signal $S = V_d \dCq$ and the system readout noise $\sigma_R = 2\sqrt{\tau_m}V_d \sigma_{\Cq}$.

\Cref{fig:tuning_drive_sweep}(c) shows the mean signal $S$ extracted from the Gaussian mixture fit for each $V_d$. A mean signal of $\SI{3.2\pm 0.2}{\femto\farad\cdot\uV}$ is reached for $V_d=\SI{15}{\uV}$  with a maximum of $\SI{3.9\pm 0.3}{\femto\farad\cdot\uV}$.
Finally in \Cref{fig:tuning_drive_sweep}(d) we extract the $V_d$ independent system readout noise $\sigma_R=2.8 \pm 0.2$ using a linear fit of $1/\sigma_{\Cq}$. 
This allows us to convert the signal in panel (c) to an SNR (right scale), with a mean SNR of $1.2\pm 0.2$ in $\SI{1}{\us}$ for $V_d=\SI{15}{\uV}$, with a maximum SNR of 1.4 $\pm 0.1$ in $\SI{1}{\us}$. The maximal SNR is in agreement with the results presented in \Cref{fig:xmpr} where the analysis was performed at a different $B_\perp$.

\begin{figure}
\includegraphics[width=\columnwidth]{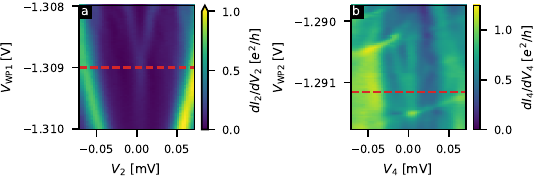}
\caption{
Transport data at junctions 2 and 4, measured in the $Z$ loop configuration before parity measurements. At junction 4, since the gates DC3L and DC4L are now open, a transport path has formed through the quantum dots that runs parallel to the wire path. This causes diagonal lines to appear in the transport scan. The data shown in \Cref{fig:zmpr} is measured at the points indicated by the red dashed lines.}
\label{fig:transport_z_loop}
\end{figure}

\paragraph{$Z$ loop parity readout.}
To demonstrate parity measurements on the bottom nanowire, we disconnect the $X$ loop by setting DC13 to depletion and connect the $Z$ loop by tuning DC3L and DC4L into the tunneling regime. This enables finite coupling between QD3 and QDL, as well as between QDL and QD4. Due to the cross capacitance from DC3L and DC4L to the top and bottom nanowire, $V_\mathrm{WP1}$ and $V_\mathrm{WP2}$ must be adjusted compared to the $X$ loop configuration.
As done prior to the $X$ loop measurement, we confirm the positions of the zero-bias peaks (ZBPs) in junctions 2 and 4 as a function of $V_\mathrm{WP1}$ and $V_\mathrm{WP2}$ by performing local differential conductance measurements (see \Cref{fig:transport_z_loop}). The observed shifts in $V_\mathrm{WP1}$ and $V_\mathrm{WP2}$ are approximately 1 mV and 2 mV, respectively.
After optimizing JTWPA performance at $f_\mathrm{QDL}$, we use the resonator connected to QDL to collect time records of $\Cq$ as a function of $V_\mathrm{QD3}$, out-of-plane magnetic field, and $V_\mathrm{QDL}$ and $V_\mathrm{QD4}$. Similar to the $X$ loop measurements, tuning the plunger voltages of the triple quantum dot allows precise control of the interferometer. To accurately quantify switching times of approximately $\SI{10}{\milli\second}$, we increase the duration of the $\Cq$ time records compared to the $X$ loop measurements to $\SI{100}{\milli\second}$. 

After identifying the quantum dot transitions with the desired tunnel couplings using a procedure similar to that described in Ref.~\onlinecite{Aghaee25}, we carry out interferometric parity measurements by varying $V_\mathrm{WP2}$ in the region where a ZBP is observed. The data presented in \Cref{fig:zmpr} qualitatively aligns with $E_{34} \approx 0$. However, as we change $V_\mathrm{WP2}$, we also find regimes displaying two h/e periodic branches at $\Ed = \pm E_{34}$, similar to the data reported in Ref.~\onlinecite{Aghaee25}, which we attribute to a sizeable $E_{34}$.

\section{Measurement software.} \label{app:software}

All measurements presented in this manuscript have been implemented using the QCoDeS~\cite{Nielsen25} open-source Python library for instrument control and data acquisition.

\vspace{1cm}
\textbf{Correspondence and requests for materials} should be addressed to Chetan Nayak (cnayak@microsoft.com).

\vspace{1cm}

$^\dagger${\small
Morteza Aghaee, Zulfi Alam, Rikke Andersen, Mariusz Andrzejczuk, Andrey Antipov, Mikhail Astafev, Lukas Avilovas, Ahmad Azizimanesh, Eric Banek, Bela Bauer, Jonathan Becker, Umesh Kumar Bhaskar, Andrea G. Boa, Srini Boddapati, Nichlaus Bohac, Jouri D.S. Bommer, Jan Borovsky, L\'eo Bourdet, Samuel Boutin, Lucas Casparis, Srivatsa Chakravarthi, Hamidreza Chalabi, Benjamin J. Chapman, Nikolaos Chatzaras, Tzu-Chiao Chien, Jason Cho, Patrick Codd, William Cole, Paul W. Cooper, Fabiano Corsetti, Ajuan Cui, Tareq El Dandachi, Celine Dinesen, Andreas Ekefj\"ard, Saeed Fallahi, Luca Galletti, Geoffrey C. Gardner, Gonzalo Leon Gonzalez, Deshan Govender, Flavio Griggio, Ruben Grigoryan, Sebastian Grijalva, Sergei Gronin, Jan Gukelberger, Marzie Hamdast, A. Ben Hamida, Esben Bork Hansen, Caroline Tynell Hansen, Sebastian Heedt, Samantha Ho, Laurens Holgaard, Kevin van Hoogdalem, John Hornibrook, Henrik Ingerslev, Lovro Ivancevic, Sherwan Jamo, Max Jantos, Thomas Jensen, Jaspreet Singh Jhoja, Jeffrey C Jones, Vidul Joshi, Konstantin V. Kalashnikov, Ray Kallaher, Rachpon Kalra, Farhad Karimi, Torsten Karzig, Seth Kimes, Evelyn King, Maren Elisabeth Kloster, Christina Knapp, Jonne V. Koski, Pasi Kostamo, Tom Laeven, Jeffrey Lai, Gijs de Lange, Thorvald W. Larsen, Kyunghoon Lee, Kongyi Li, Guangze Li, Shuang Liang, Tyler Lindemann, Matthew Looij, Marijn Lucas, Roman Lutchyn, Morten Hannibal Madsen, Nasiari Madulid, Michael J. Manfra, Laveena Manjunath, Signe Markussen, Esteban Martinez, Marco Mattila, J. R. Mattinson, R. P. G. McNeil, Alba Pérez Millan, Ryan V. Mishmash, Sarang Mittal, Christian M{\o}llgaard, M. W. A. de Moor, Eduardo Puchol Morejon, Trevor Morgan, George Moussa, B.P. Nabar, Anirudh Narla, Chetan Nayak, Jens Hedegaard Nielsen, William Hvidtfelt Padkær Nielsen, Frédéric Nolet, Michael J. Nystrom, Eoin O'Farrell, Thomas A. Ohki, Keita Otani, Camille Papon, Karl D Petersson, Luca Petit, Dima Pikulin, Mohana Rajpalke, Alejandro Alcaraz Ramirez, David Razmadze, Yuan Ren, Ivan Sadovskyy, Lauri Sainiemi, Juan Carlos Estrada Salda\~na, Irene Sanlorenzo, Tatiane Pereira dos Santos, Simon Schaal, John Schack, Emma R. Schmidgall, Christina Sfetsou, Cristina Sfiligoj, Sarat Sinha, Patrick Sohr, Thomas L. S{\o}rensen, Kasper Spiegelhauer, Toma\v{s} Stankevi\'c, Lieuwe J. Stek, Patrick Str{\o}m-Hansen, Henri J. Suominen, Judith Suter, Samuel M. L. Teicher, Raj Tholapi, Mason Thomas, D.W. Tom, Emily Toomey, Joshua Tracy, Michelle Turley, Matthew D. Turner, Shivendra Upadhyay, Ivan Urban, Dmitrii V. Viazmitinov, Anna Wulff Viazmitinova, Beatriz Viegas, Dominik J. Vogel, John Watson, Alex Webster, Joseph Weston, Timothy Williamson, Georg W. Winkler, David J. van Woerkom, Brian Paquelet Wuetz, Chung-Kai Yang, Shang-Jyun (Richard) Yu, Emrah Yucelen, Jes\'us Herranz Zamorano, Roland Zeisel, Guoji Zheng, A.M. Zimmerman
}

\vspace{0.5 cm}

\textbf{Affiliation.} All authors contributed to this manuscript as employees of Microsoft Quantum.

\bibliography{xmpr}

\begin{thebibliography}{83}
\bibitem [{Zti()}]{Ztimescale_footnote}%
  \BibitemOpen
  \href@noop {} {}\bibinfo {note} {Note that the kurtosis-based selection of
  time records together with the finite \SI{100}{\milli\second} length of time
  records may bias towards shorter dwell times. Instances of the distribution
  with long dwell time realizations may show a significant asymmetry between
  the two states of the RTS due to a small number of switches. This leads to
  the white spots in the kurtosis map of \Cref{fig:zmpr}(b) and excluding these
  points from the analysis by applying the kurtosis threshold may lead to
  underestimating $\tau_\Z$.}\BibitemShut {Stop}%
\bibitem [{SNR()}]{SNR_footnote}%
  \BibitemOpen
  \href@noop {} {}\bibinfo {note} {Specifically, we extract an SNR of 1.9 in
  $\SI{2}{\micro\second}$ from the Gaussian fits in \Cref{fig:xmpr}(e). We
  rescale this by $\sqrt{2}$ to extrapolate to $\SI{1}{\micro\second}$. Note
  that the SNR from the autocorrelation function $\dCq / 2\sigma_{\Cq} = 1.66$
  is slightly different, likely due to the GMM fits being sensitive to
  additional noise components with finite correlation time.}\BibitemShut
  {Stop}%
\end{thebibliography}

\end{document}